\documentclass[journal]{IEEEtran}
\ifCLASSOPTIONcompsoc
  \usepackage[nocompress]{cite}
\else
  \usepackage{cite}
\fi
\ifCLASSINFOpdf
\else
\fi
\usepackage{afterpage}
\usepackage{threeparttable}
\usepackage [english]{babel}
\usepackage [autostyle, english = american]{csquotes}
\MakeOuterQuote{"}
\usepackage{amsmath}

\usepackage{graphicx}
\usepackage{subfig}
\usepackage{comment}
\usepackage[dvipsnames]{xcolor}
\graphicspath{ {./images/} }
\usepackage{hyperref}
\usepackage[switch]{lineno}

\begin{document}
%
\title{Cross-Subject Emotion Recognition with Sparsely-Labeled Peripheral Physiological Data Using SHAP-Explained Tree Ensembles}

\author{Feng~Zhou,
        Tao Chen,~\IEEEmembership{Member, IEEE},
        Baiying  Lei,~\IEEEmembership{Senior Member, IEEE}
\IEEEcompsocitemizethanks{\IEEEcompsocthanksitem F. Zhou is with Department of Industrial and Manufacturing Systems Engineering, University of Michigan-Dearborn, Dearborn, Michigan 48128, USA
\protect\\
E-mail: fezhou@umich.edu
\IEEEcompsocthanksitem T. Chen is with School of Information Science and Technology, Fudan University, Shanghai, 200433, China.}
\thanks{Manuscript received xxx, 2020; revised xxxx, 2020.}}

%
%

\markboth{IEEE Transactions on Human-Machine Systems,~Vol.~xx, No.~xx, October~2022}%
{Zhou \MakeLowercase{\textit{et al.}}: Cross-Subject Emotion Recognition}
%



\IEEEtitleabstractindextext{%
\begin{abstract}
There are still many challenges of emotion recognition using physiological data despite the substantial progress made recently. In this paper, we attempted to address two major challenges. First, in order to deal with the sparsely-labeled physiological data, we first decomposed the raw physiological data using signal spectrum analysis, based on which we extracted both complexity and energy features. Such a procedure helped reduce noise and improve feature extraction effectiveness. Second, in order to improve the explainability of the machine learning models in emotion recognition with physiological data, we proposed Light Gradient Boosting Machine (LightGBM) and SHapley Additive exPlanations (SHAP) for emotion prediction and model explanation, respectively. The LightGBM model outperformed the eXtreme Gradient Boosting (XGBoost) model on the public Database for Emotion Analysis using Physiological signals (DEAP) with f1-scores of 0.814, 0.823, and 0.860 for binary classification of valence, arousal, and liking, respectively,  with cross-subject validation using eight peripheral physiological signals. Furthermore, the SHAP model was able to identify the most important features in emotion recognition, and revealed the relationships between the predictor variables and the response variables in terms of their main effects and interaction effects. Therefore, the results of the proposed model not only had good performance using peripheral physiological data, but also gave more insights into the underlying mechanisms in recognizing emotions.  
\end{abstract}

\begin{IEEEkeywords}
Emotion Recognition, Peripheral Physiological Data, Tree Ensembles, Explainability, SHAP.
\end{IEEEkeywords}}

\maketitle

\IEEEdisplaynontitleabstractindextext

%
\IEEEpeerreviewmaketitle

\section{Introduction}

%
%
%
%
\IEEEPARstart{A}{ffect} is one essential element in the human-computer interaction process \cite{picard2000affective} and can be included as a design parameter to optimize task performance and user experience in various areas \cite{zhou2014prospect}. For example, the Yerkes-Dodson law identified an inverted-U shape relationship between task performance and arousal \cite{yerkes1908relation}, participants' positive valence was found to improve their takeover performance in automated driving \cite{du2020examining}, and recognizing and understanding emotions were able to help improve human-robot interaction \cite{chen2017dynamic}. 

In order to support affective computing, one basic question is how to recognize emotions during the interaction process \cite{zhou2011affect}. Researchers explored different types of data to recognize emotions, including self-reported data, behavioral measures (e.g., facial expressions \cite{zhou2020fine} and conversational data \cite{sun2019design}), and physiological measures (e.g., facial electromyography (EMG), skin conductance response (SCR), electrocardiogram (ECG), electroencephalography (EEG)) \cite{zhou2011affect,zhou2014emotion,zhou2020driver}. Although self-reported methods are easy to operate, they suffer from recall and selective reporting biases \cite{stone1994ecological} and interference with the interaction process \cite{zhou2020driver}. The progress on emotion recognition using facial expressions and speeches is promising due to the recent advancement in deep learning (e.g., \cite{zhao2019speech,zhou2020fine}). Nevertheless, such affective displays are subject to social display rules and the expressions might not be consistent with their real emotions \cite{picard2000affective,frantzidis2010classification}. By contrast, physiological measures are less vulnerable to voluntary control and allow the system to evaluate users' emotions in real time continuously \cite{zhou2011affect}. The continuous fashion in recognizing emotion is consistent with human perception on emotions in social interaction \cite{schiano2004categorical}, and the real-time characterization is preferable for human-computer interaction systems that aim to fulfill users' emotional needs by addressing their emotions in real time \cite{picard2000affective,zhou2013affective,zhou2020emotional}. Therefore, many studies have successfully built machine learning models based on physiological measures (e.g., \cite{kim2008emotion,zhou2011affect,walter2013transsituational,6846297, wen2014emotion,ma2019emotion,romeo2019multiple,lan2019domain,li2020multisource,chen2021personal}). 

Correspondingly, an increasing number of datasets (e.g.,\cite{koelstra2011deap,soleymani2011multimodal,zheng2018emotionmeter}) have been created to train models to recognize emotions automatically using physiological data. However, one of the biggest challenges in training models is to label the physiological data. Unlike other databases that have one label for each sample (e.g., facial expression images in AffectNet \cite{mollahosseini2017affectnet}), physiological data were continuously recorded when the participant was involved in an emotional episode. Thus, it is extremely time-consuming to label the data frame by frame, which is impossible \cite{romeo2019multiple}. Many datasets only had an overall label for a continuous segment of physiological data recorded in the whole emotional event, such as the Database for Emotion Analysis using Physiological signals (DEAP)  \cite{koelstra2011deap}, which was only sparsely labeled using self-reports or expert observation. Here sparse labelling \cite{romeo2019multiple} was used to indicate that there was only one or a few static labels annotating a segment of physiological data, which was supposed to correspond to dynamic changes in emotional states (i.e., continuous labels).
Self-reported labelling was often conducted once (due to possible interference in the interaction process) after the participants finished the interaction with the stimuli, such as video or image viewing \cite{zhou2011affect,zhou2014emotion,koelstra2011deap} and game playing \cite{yannakakis2008entertainment}. Labeling by expert observation can potentially increase the number of the labels in a segment of physiological data, and multiple experts should be involved in order to increase reliability of the labels (e.g., \cite{li2017automatic}). However, both of these methods only label the most salient one within the whole emotional episode rather than the dynamic changes of emotions of the participants. Such labels tend to be noisy when the most salient period only accounts for a small period of the whole labeled episode \cite{romeo2019multiple}.

Another challenge for many machine learning models in emotion recognition is that they are black-box models (especially for deep learning models) and the knowledge learned by them is hidden. For one thing, without revealing the knowledge learned, the acceptance and trust in such models can be jeopardized \cite{doshi2017towards}, especially in high risk areas such as medicine \cite{lundberg2018explainable,yan2020interpretable} and transportation \cite {zhou2020driver,AYOUB2021102}.
For another, when such knowledge is revealed, it can help distinguish between different emotions. For instance, Weitz et al. \cite{weitz2019deep} used layer-wise relevance propagation and local interpretable model-agnostic explanations (LIME) to understand the decision made by a deep neural network, so that they were able to distinguish facial expressions caused by pain and disgust in order for the nurses to give better care for their patients. 

Given the above challenges, the objective of this paper is to recognize emotions from sparsely-labeled peripheral physiological data with explanations. First, we applied singular spectrum analysis (SSA) to decompose the physiological data into individual components, and extracted three types of features from selected components, including sample entropy, fuzzy entropy, and energy. SSA has been widely applied in time series analysis and it can decompose the raw signal into a small number of independent and interpretable components (e.g., periodic components, slow-varying components, and noise components) \cite{elsner2013singular}, and it has been proved to improve the performance of classification models built on physiological data by removing random noise and data reconstruction at the same time. Entropy measures have also been widely used for feature extraction of physiological measures for classification purposes with good performance. For example, Lu et al. \cite{lu2020entropy} extracted entropy features from SSA components of various physiological measures, including EEG, EMG, and ECG, across multiple classification scenarios and the performance was much better than extracting features directly from the raw physiological signals. Particularly, sample entropy was used to predict arrhythmic risk using ECG signals \cite{rajesh2017classification} and to recognize emotions using EEG \cite{lu2020dynamic}, fuzzy entropy was used to detect epileptic seizure with EEG signals \cite{kumar2014epileptic}, sample entropy and fuzzy entropy were used to predict eye states using EEG signals, physical actions using EMG signals, and heart states using ECG signals \cite{lu2020entropy}. The energy feature was similar to the power spectral density, which was also used in emotion recognition \cite{jenke2014feature}.  
Second, in order to build explainable machine learning models, we applied tree-based ensemble models, including eXtreme Gradient Boosting (XGBoost) \cite{chen2016xgboost} and Light Gradient Boosting Machine (LightGBM) \cite{ke2017lightgbm} and SHapley Additive exPlanations (SHAP) \cite{lundberg2018explainable,lundberg2020local}. The entropy and energy features extracted from SSA components were used to train the XGBoost and LightGBM models. The LightGBM model was found to perform better than the XGBoost model, and we used SHAP to explain the LightGBM model by identifying the most significant features and their main and interaction effects. Such ensemble methods have proved to perform better than other traditional machine learning models, such as support vector machines (SVMs), due to the fact that ensemble learning overcomes the shortcomings of single classifiers by combining dissimilar classifiers \cite{9652403,zhou2022predicting,zhou2021using,ayoub2021modeling}. SHAP is based on the game theory to explain the output of the LightGBM model by examining feature contributions in a coalition using the classic Shapley values \cite{shapley1953contributions} and it has three desirable mathematical properties, i.e., local accuracy, missingness, and consistency, which provide consistent and locally accurate explanations \cite{lundberg2017unified,lundberg2018explainable}. However, the computational requirements make it difficult to apply to many machine learning models with a large number of samples, especially for deep learning models. Lundberg et al. \cite{lundberg2020local} proposed an algorithm to explain tree-based models efficiently using SHAP. Hence, by making use of SHAP and LightGBM, we were able to understand the patterns captured by the LightGBM model in recognizing emotions, by  helping select and explain the most important SSA components of their main effects and interaction effects. In addition, both the machine learning models were validated using a leave-one-subject-out (LOSO) scheme (i.e., user-independent), thus the hidden knowledge revealed tended to be more generalizable.

In summary, the contributions of this paper are described as follows:
\begin{itemize}
  \item We decomposed the peripheral physiological data into individual SSA components and extracted both entropy and energy features to deal with the sparse label issue. Such a method explores a wide range of feature space to improve emotion representation across a wide range of signals while removing random noise at the same time.
  \item We built explainable machine learning models using tree-based ensemble machine learning models (i.e., XGBoost and LightGBM) for emotion recognition, and used SHAP to identify the most important SSA components to improve our results with LOSO cross-validation. Such a feature selection method provides useful information and implications on the parameter selection in SSA decomposition.
  \item We identified the main effects and interaction effects that showed the hidden relationships between the emotion representations and the most important features identified. Such insights were  used to improve the performance of the proposed models and have the potential to pave the way for future applications with emotion recognition in different domains.
\end{itemize}

\section{Related Work}
\subsection{Emotion Recognition Using Physiological Data}
Emotions can be recognized from physiological data collected from the central nervous system and the peripheral nervous system. The most frequently used signal from the central nervous system is EEG (e.g., \cite{koelstra2011deap,yin2017recognition,ma2019emotion}) while there are a large number of signals that can be collected from the peripheral nervous system, such as SCR, EMG, electrooculargram (EOG), and ECG (e.g., \cite{picard2001toward,zhou2011affect,romeo2019multiple}). In order to recognize emotions, various machine learning models were applied. Picard et al. \cite{picard2001toward} used SCR, EMG, respiration data, and blood volume pulse of a single participant to predict eight emotions with Fisher projection with 81\%  accuracy. Liu et al. \cite{liu2008physiology} used ECG, photoplethysmogram (PPG), phonocardiogram, SCR, and facial EMG to recognize three emotions using SVMs, which obtained 82.9\% accuracy in recognizing emotions from autistic children. Many researchers made use of the public DEAP dataset for binary classification. For example, Yin et al. \cite{yin2017recognition} applied an ensemble classifier using stacked autoencoder and extracted 425 features from EEG, EOG, GSR, PPG, EMG, respiration, and skin temperature and obtained f1-scores of 77.98\%  for arousal recognition and of 79.50\% for valence recognition. Lin et al. \cite{lin2017deep} proposed a convolutional neural network (CNN) by transforming EEG data into image data with time and frequency information, and other hand-crafted features from peripheral physiological data with 87.30\% accuracy (f1-score: 72.84\%) for arousal recognition and 85.5\% accuracy (f1-score: 80.06\%) for valence recognition. Ma et al. \cite{ma2019emotion} proposed a deep learning model named multimodal residual long-short-term memory (LSTM) with an accuracy of 92.87\% and 92.30\% for binary arousal and valence classification. Despite the good performance of the deep learning models, these models were user-dependent, i.e., models trained using data from one specific participant, and thus were less likely to generalize across participants due to potential individual differences.  

Other researchers built user-independent models by mixing data from multiple participants with $k$-fold cross validation. For example, Nasoz et al. \cite{nasoz2004emotion} used SCR, skin temperature, and heart rate to recognize six emotions from 29 participants and obtained the best accuracy of 83\% using feedforward neural networks trained by Marquardt back propagation. Frantzidis \cite{frantzidis2010classification} applied a decision tree model with a Mahalanobis distance classifier to recognize four emotions with EEG and SCR data from 28 participants with 78\% accuracy. Zhou et al. \cite{zhou2011affect} proposed machine learning models using rough set to recognize seven discrete emotions elicited by visual stimuli from 42 participants. They proposed gender-specific, culture-specific, and general models and found that models built from female participants (i.e., gender-specific) had the best performance of 78.0\% while general participant-independent models only had 63.5\% in terms of f1-scores. Wen et al. \cite{wen2014emotion} used a multi-variant correlation method of various physiological signals, including blood oxygen saturation, SCR, and heart rate to identify useful features, based on which a random forest model was proposed to recognize four emotions with 74\% accuracy.  Ringeval et al. \cite{ringeval2015prediction} proposed an LSTM regression model by combining audio, visual (facial expressions) and  physiological (ECG, SCR) data from the Remote Collaborative and Affective Interactions (RECOLA) dataset  \cite{ringeval2013introducing}, and the concordance correlation coefficients were 0.804 for arousal and 0.528 for valence. Li et al. \cite{li2016analysis} included group-based individual response specificity in building emotion recognition models with ECG, SCR, and PPG and their performance was better (f1-score: 0.75) than a general model (f1-score: 0.67) without such information. Hassan et al. \cite{hassan2019human} proposed a deep belief network and an SVM model and achieved 89.53\% accuracy on the DEAP dataset for recognizing 5 discrete emotions using EMG, PPG and SCR data. 

The user-independent models with $k$-fold cross-validation mentioned above still assume that the samples are independent and identically distributed. However, the samples from one participant are actually correlated biologically and temporally \cite{dehghani2019subject}. Another type of validation strategy that tends to have better generalization is LOSO cross-validation. In this process, one subject is randomly selected for testing while the rest ones are used for training the model and this procedure is repeated until all the subjects are used for testing \cite{koul2018cross}. This validation strategy is especially useful for the model to predict emotions of new individuals \cite{saeb2017need}. 
Romeo et al. \cite{romeo2019multiple} proposed a multiple instance learning strategy with SVMs to recognize emotions using physiological measures from the DEAP dataset. The multiple instance learning treated the data as negative if all the instances were negative and positive if at least one of them was positive. They stated that using the LOSO cross-validation, the results were not satisfactory on the DEAP dataset and thus were not reported. On their own dataset, they had the best binary classification accuracy of 69.1\% for arousal and 68.0\% for valence. Kandemir et al. \cite{kandemir2014multi} proposed a multi-task multi-view learning model with SVMs. They reported their LOSO cross-validation performance on the DEAP dataset using both EEG and peripheral physiological measures was 60\% (accuracy) and 56\% (f1-score) for valence, 58\% (accuracy) and 53\% (f1-score) for arousal, and 67\%(accuracy) and 51\% (f1-score) for liking. Zhong et al. \cite{zhong2017emotion} proposed a temporal information preserving framework to identify the relations between physiological data and emotions over time and by combining both physiological and facial expression data, they obtained the best accuracy for valence and arousal as 70.53\% and 73.53\% on the MAHNOB-HCI dataset using LOSO cross-validation\cite{soleymani2011multimodal}. 
Gupta et al. \cite{gupta2019cross} proposed a flexible analytic wavelet transform to explore channel specific EEG signals in the forms of different sub-band forms for cross-subject emotion recognition with random forest and support vector machines. Another way to help improve the performance of LOSO is to minimize individual differences. Chen et al. \cite{chen2021personal} proposed a personal-Zscore feature processing method, which was found to improve the representation ability of emotion across different subjects, which improved emotion recognition on the SEED dataset. Li et al. \cite{li2020multisource} proposed a transfer learning method that reduced the EEG differences between the target (i.e., the new person's data in the testing set) and each source (i.e., the existing participants' data in the training set) that improved three-category classification accuracy on SEED dataset by 12.72\%.  Lan et al. \cite{lan2019domain} proposed a domain adaptation technique to reduce cross-subject variance and they improved the accuracy from the baseline by 7.25\%-13.40\% on the DEAP and SEED datasets.

\subsection{Explainable Machine Learning Models} 
We have witnessed great successes of machine learning models in various areas as they can capture the hidden domain knowledge and make predictions for unseen data. Nevertheless, these models can be less acceptable and trusted during the deployment stage without showing the captured domain knowledge, i.e., explainability \cite{doshi2017towards}. Simpler machine learning models, such as linear regression models and decision trees are explainable. However, the majority of the other machine learning models (e.g., SVMs, LSTM, deep belief networks, CNNs) reviewed above in emotion recognition are black-box models and thus are not explainable. Although the importance of explainability in decision making with high risks is greater (e.g., medicine \cite{lundberg2018explainable,yan2020interpretable}, transportation \cite {zhou2020driver}, and misinformation \cite{ayoub2021combat}), it is still useful in different affective computing applications (e.g.,\cite{weitz2019deep}). In addition, the choice between two types of models, i.e., simple, explainable models and complicated, black-box models tends to be easily made in emotion recognition as simpler models are not able to have good performance. Therefore, the only way to understand the captured domain knowledge is to apply post-hoc explainability to black-box models \cite{murdoch2019definitions}. Very few studies attempted to explain models in emotion recognition in the literature. LIME was used to explain a CNN model to distinguish between pain and disgust facial expressions that helped nursing staff identify pain episodes of patients by monitoring their facial expressions \cite{weitz2019deep}. 

\section{Methods} 
\begin{figure*} [htbp]
\centering
\includegraphics[width=14cm]{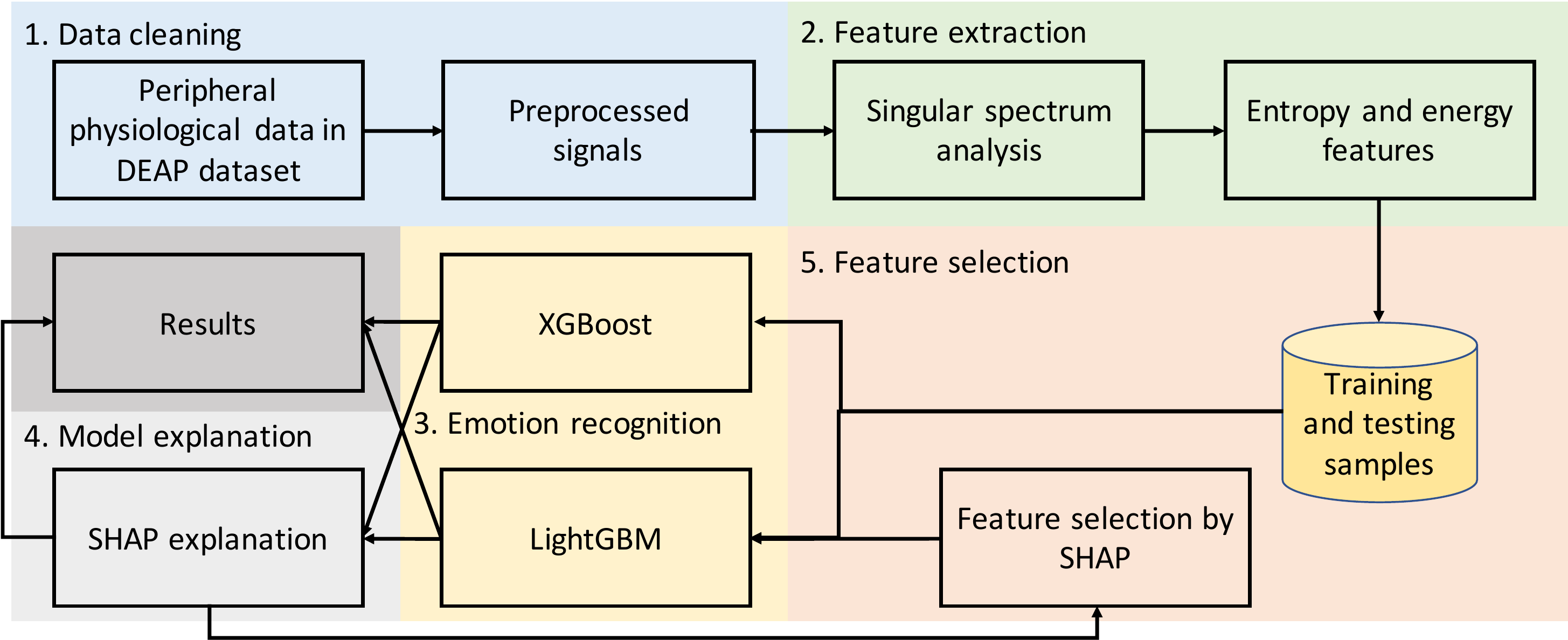}
\caption{The major steps involved in the proposed method, including 1) data cleaning, 2) feature extraction, 3) emotion recognition, 4) model explanation, and 5) feature selection.}
\label{fig:method}
\end{figure*}
The proposed method has five major steps (see Fig. \ref{fig:method}), including 1) data cleaning, 2) feature extraction, 3) emotion recognition, 4) model explanation,  and 5) feature selection as described below. Note the third step and the fourth step were conducted twice before and after the fifth step in order to select the optimal features for best performance. 

\subsection{Dataset and Data Preprocessing}
In this paper, the DEAP dataset \cite{koelstra2011deap} was used to demonstrate the proposed method with eight peripheral physiological signals, including horizontal EOG (i.e., hEOG $=$ hEOG\_1 $-$ hEOG\_2), vertical EOG (i.e., vEOG $=$ vEOG\_1 $-$ vEOG\_2), zygomaticus major EMG (i.e., zEMG $=$ zEMG\_1 $-$ zEMG\_2), trapezius EMG (i.e., tEMG $=$ tEMG\_1 $-$ tEMG\_2), SCR, PPG, respiration (Resp), and skin temperature (Temp). The data were downsampled from 512 Hz to 128 Hz, which was adequate and efficient for processing the peripheral physiological signals and this has been widely used in previous studies \cite{piho2018mutual, romeo2019multiple}. The locations of each electrode during the data collection process are shown in Table \ref{tab:placement}.

\begin{table}[]
\centering
\caption{Placement of electrodes in collecting physiological data in the DEAP dataset \cite{koelstra2011deap}.}
\label{tab:placement}
\begin{tabular}{ll}
\hline
Signal         & Electrode placement                  \\ \hline
hEOG\_1        & to the left of left eye              \\
hEOG\_2        & to the right of right eye            \\
vEOG\_1        & above right eye                      \\
vEOG\_2        & below right eye                      \\
zEMG\_1        & $ +/-$ 1cm from left corner of mouth \\
zEMG\_2        & $+/-$ 1cm below zEMG\_1              \\
tEMG\_1        & left shoulder blade                  \\
tEMG\_2        & $+/-$ 1cm below tEMG\_1              \\
GSR            & left middle and ring fingers          \\
Respiration    & respiration belt around chest        \\
PPG & left thumb                           \\
Temperature    & left pinky                           \\ \hline
\end{tabular}%
\end{table}
A number of 32 participants watched 40 music videos with 60 seconds long and rated on valence, arousal, and liking. Between each trial, there were five seconds of fixations on screen before the trial and three seconds fixations on screen after the playback of each music video. In this study, we included data of the first three seconds with fixations on the screen as the baseline and of the rest 60 seconds of video watching as the usable signals. There was a short break after watching 20 videos and the estimated length of the experiment was about 90 minutes. Thus, there were $40 \times 32 = 1280$ samples in total. The labels used in this paper included valence, arousal, and liking. Following previous studies, they were binarized by thresholding at 5 from the 9-point Likert scale (e.g., \cite{kandemir2014multi, romeo2019multiple}). 

First, the samples were smoothed with a moving average filter with a span of 64 data points (i.e., 0.5 second since the signals had a sampling rate of 128 Hz) for SCR and Temp and of 5 data points for other signals. Second, the signals were baseline corrected, i.e., the 60 seconds of the usable data were subtracted from the mean of the baseline. Third, each sample was normalized by subtracting the mean and divided by the standard deviation. Last, each sample was detrended quadratically with Matlab 2019b (Mathworks, Natick, MA), except Temp and SCR. Ledalab was used to decompose SCR into a phasic component and a tonic component \cite{benedek2010decomposition}.

\subsection{Signal Decomposition with SSA}
The decomposition of the physiological signals using SSA was as follows \cite{elsner2013singular}. Each sample was represented as a time series signal $X_N=[x_1,x_2,...,x_N]$ of length $N = 60 \times 128 = 7680$. The first step of SSA is to transform $X_N$ into a multi-dimensional Hankel matrix $Y = [Y_1,Y_2,...,Y_K]$ with $Y_i = [x_i,x_{i+1},...,x_{i+L-1}]^T$, where $K=N-L+1$. This process was named as embedding and its parameter is the window length $L$ and $2\leq L\leq N$. In this research, $L$ was set at 12 based on previous studies \cite{lu2020entropy}. Then, singular value decomposition was applied to the matrix, $S=YY^T$. Assuming $\lambda_1\geq\lambda_2\geq ... \geq\lambda_L$ were the eigenvalues of $S$ and $U_1,U_2,...,U_L$ was the orthonormal basis of the eigenvectors of $S$. By setting $V_i = Y^TU_i/sqrt(\lambda_i)$ and the rank of $Y$ as $d = max(i,for \lambda_i>0)$, $Y$ was represented as 
\begin{equation} \label{1}
Y = Y_1+Y_2+,...,+Y_d,
\end{equation}
where $Y_i=sqrt(\lambda_i)U_iV_i^T$. 

In the reconstruction stage, it had two steps, including eigentriple grouping and diagonal averaging. First, the matrices $Y_1,Y_2,...,Y_d$ were divided into $m$ disjoint groups, i.e., $Y = Y_{I1}+,...,Y_{Im}$, where $I = \{i_1,...,i_p\}$ is a group of indices selected from $ {1,...,d}$ so that $Y_I = Y_{i1}+,...,+Y_{ip}$. Then, diagonal averaging converted the matrix $Y_{Ik}$, where $1\leq k \leq m$, into a time series by averaging $x_{ij}$ along the diagonal of the matrix so that $i+j=constant$. For the $k$-th reconstructed time series, $Y^{k} = [y_1^k,...,y_N^k]$ and the original time series $X_N=[x_1,x_2,...,x_N]$ were reconstructed as a sum as follows:
\begin{equation} \label{1}
x_{n}=\sum_{k=1}^{m}y_n^k,n=1,...,N
\end{equation}

\begin{figure} [bpt]
	\centering
	\subfloat[\label{fig:PPG1}]{\includegraphics[width=0.95\linewidth]{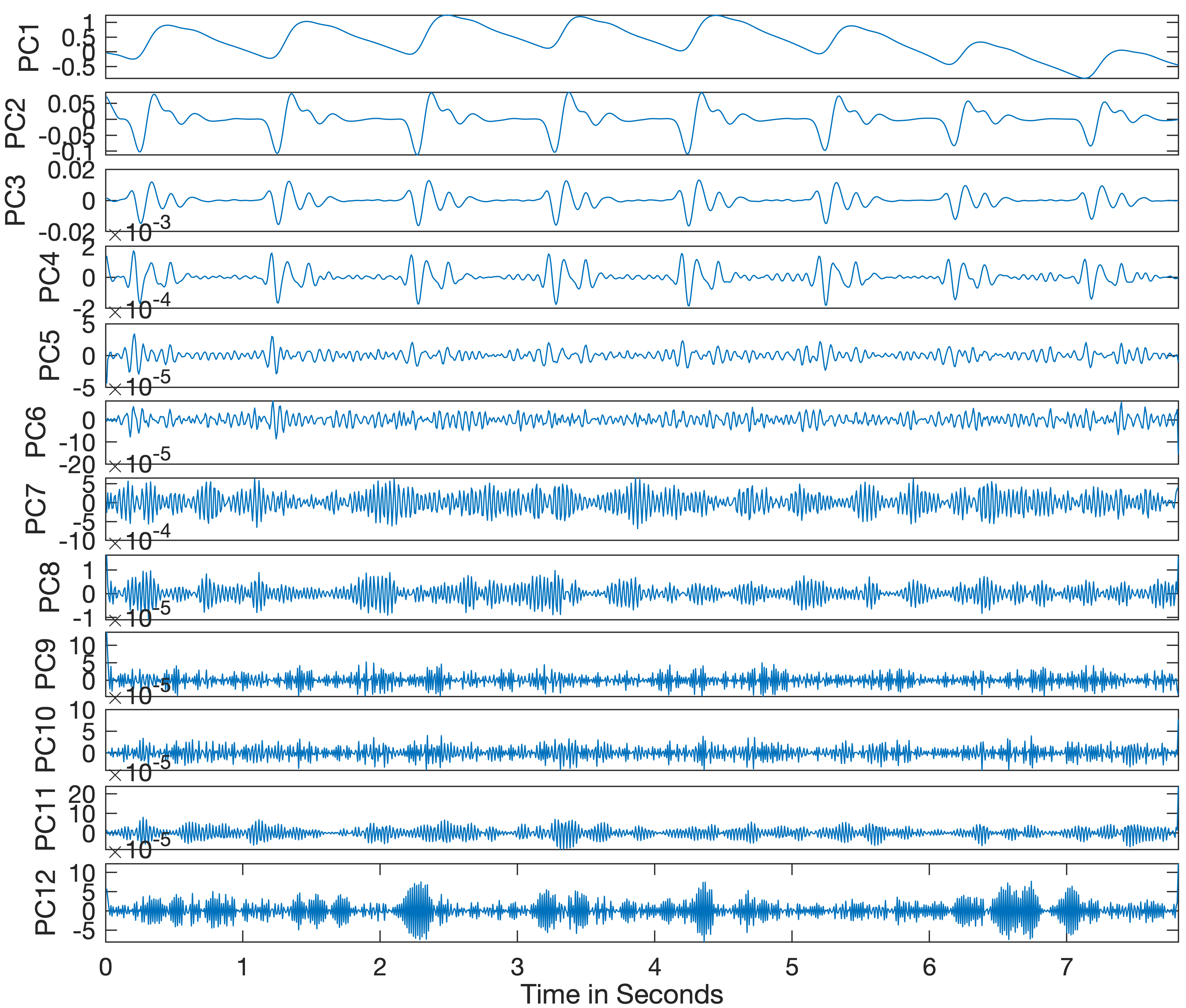}}
	\vspace{0pt}
	\subfloat[\label{fig:PPG2}]{\includegraphics[width=.95\linewidth]{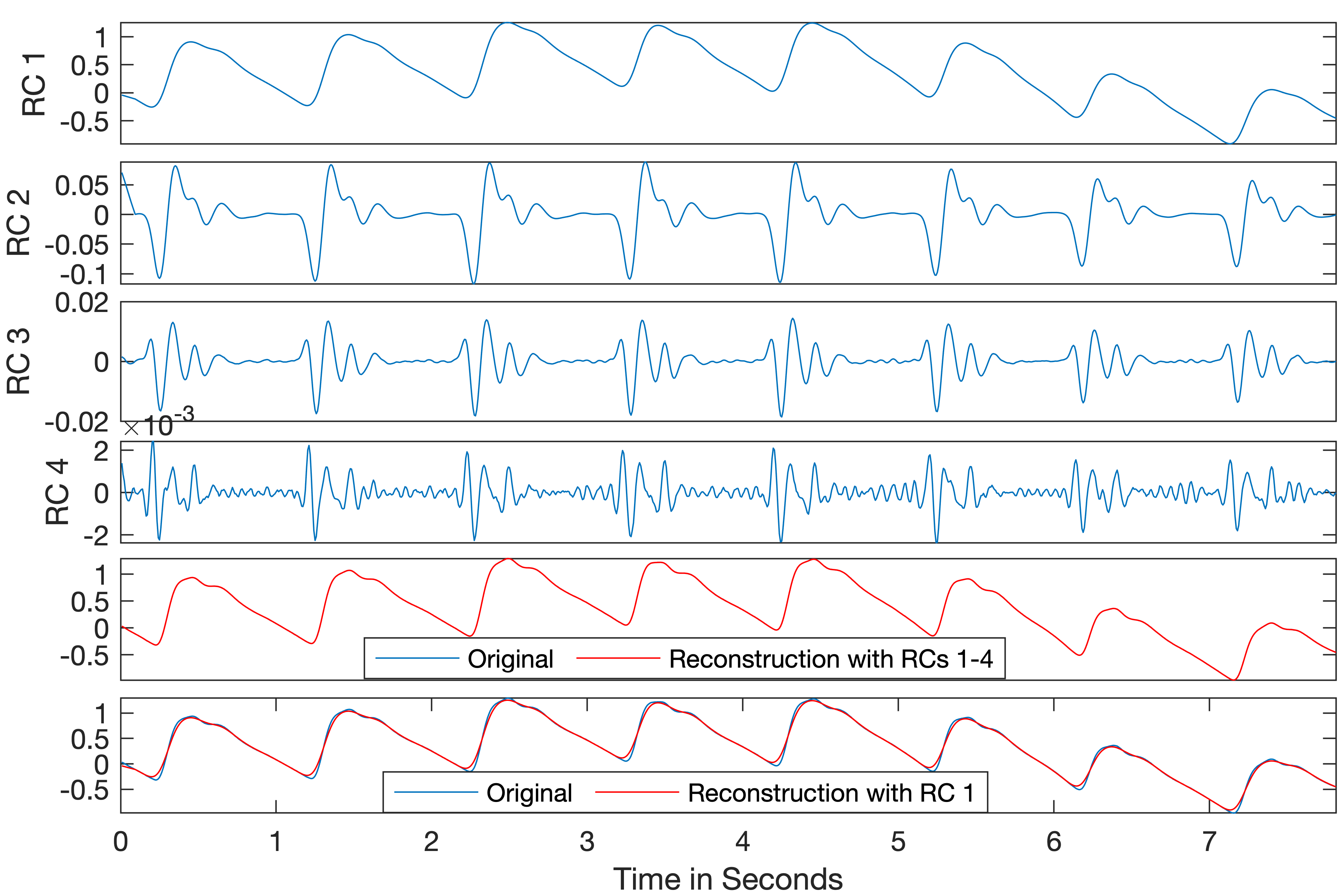}}
    \caption{An example of decomposed PPG signals using SSA: (a) The 12 principal components (i.e., PC in the figure) derived in one PPG sample and the first four were used for reconstruction while the rest ones were considered as noise; (b) PPG reconstructed components (i.e., RC in the figure) and the comparison between the reconstructed PPG signal (using the RCs in the top 4 panels) and the original PPG signal (the fifth panel from the top). Note the reconstructed PPG signal with RCs 1-4 was so close to the original PPG signal that there were no visible differences due to the scale of the figure. We also added the sixth panel (from the top) that showed the comparison between the reconstructed PPG signal with only RC 1 and the original PPG signal.}\label{fig:SSA}
\end{figure}

In this research, we used the pymssa package in Python (\url{https://github.com/kieferk/pymssa}) for SSA implementation and only kept a selected number of components to remove the noise, which proved to be effective for feature extraction \cite{lu2020entropy}. Assuming $X_N = S_N + R_N$, we only selected $I_k = \{1,...,s\}, s < d$ in order to remove the noise component $R_N$. We used the formula of singular value hard thresholding to remove the noise components, which proved to be able to recover matrices with bounded nuclear forms with the best possible asymptotic mean squared error \cite{6846297}. 
For example, Fig. \ref{fig:SSA} (a) shows the 12 principal  components of a PPG signal and the formula of singular value hard thresholding kept the first four components for reconstruction and Fig. \ref{fig:SSA} (b) shows the four reconstructed components (the top four panels) and the comparison between the original signal and the reconstructed one (the fifth panel) with four reconstructed components. Similarly, the first two, two, one, three, two, two, and one component(s) were kept for hEOG, vEOG, zEMG, tEMG, SCR, Resp, and Temp, respectively, for reconstruction and feature extraction. Note the SCR signal used for SSA decomposition was the phasic component of the SCR data and the tonic component was used as the second component for feature extraction. Thus, a total number of 17 SSA components were extracted and we calculated their sample entropy, fuzzy entropy, and energy, which resulted in a total number of 51 features.

\subsection{Feature Extraction}
Three features were extracted for each component derived in the previous step, including sample entropy, fuzzy entropy, and energy. Sample entropy and fuzzy entropy are complexity measures. Sample entropy is based on approximate entropy, which is used to measure the degree of regularity and classify complex systems of both deterministic chaotic and stochastic processes  \cite{pincus1991approximate}. However, it is heavily dependent on the data length (at least 1000 data points) and can lead to inconsistency results for short and noisy data \cite{richman2000physiological}. Sample entropy is an improved version of approximate entropy by removing data length dependency and was widely used as features extracted from physiological signals for classification purposes \cite{richman2000physiological,lu2020dynamic,lu2020entropy}. Fuzzy entropy makes fuzzy measurement of signals by introducing a fuzzy membership function to avoid abrupt changes, which can be especially useful for characterizing EMG signals \cite{chen2007characterization}. 

Given an SSA component, $Y^{k} = [y_1^k,...,y_N^k]$, a non-negative integer $m\leq N$ as the embedding dimension, and a positive real number $r$ as the tolerance, we calculated 
the distance $d[Y_m^k(i),Y_m^k(j)]= max_{k=1,2,...,m,i\neq j}|Y_m^k(i+k-1)-Y_m^k(j+k-1)|$, and sample entropy was calculated as \cite{richman2000physiological}
\begin{equation} \label{SampEn}
SampEn =  -log \frac{A}{B}, 
\end{equation}
where $A$ was the number of template vector pairs that satisfied $d[Y_{m+1}^k(i),Y_{m+1}^k(j)]<r$ and $B$ was the number of template vector pairs that satisfied $d[Y_{m}^k(i),Y_{m}^k(j)]<r$.

In calculating fuzzy entropy, $Y_m^k(i)$ was defined as  $[y_i^k,y_{i+1}^k,...,y_{i+m-1}^k]-y^0_i$, where $y^0_i$ was the baseline defined as $ \frac{1}{m}\sum_{j=0}^{m-1}y_{i+j}^k$. The distance $d[Y_m^k(i),Y_m^k(j)]= max_{k=0,1,...,m-1,i\neq j}|Y_m^k(i+k-1)-Y_m^k(j+k-1)|$. Given a non-negative integer $n$, the similarity degree $D_m^{k}(i,j)$ between $Y_m^k(i)$ and $Y_m^k(j)$ was computed as $D_m^{k}(i,j)=exp(-(d[Y_m^k(i),Y_m^k(j)])^n/r)$. Then we used the following to calculate fuzzy entropy \cite{chen2007characterization}
\begin{equation} \label{FuzzyEn}
FuzzyEn=ln\Phi^{m}(n,r)- ln\Phi^{m+1}(n,r),
\end{equation}
where $ln\Phi^{m}(n,r) = \frac{1}{N-m}\sum_{i=1}^{N-m}(\frac{1}{N-m-1}\sum_{j=1}^{N-m}D_m^{k}(i,j))$. 
According to \cite{lu2020entropy}, we used embedding dimension, $m=2$ and tolerance, $r=0.15$ in calculating entropy ($n=2$ for fuzzy entropy) measures for each SSA component derived from the physiological data. In addition, energy was also calculated for the $k$-th SSA component of a physiological signal using
\begin{equation} \label{Energy}
E=\frac{1}{N}\sum_{i=1}^{N}(y_i^k)^2, 
\end{equation}
where $N=7680$ was the number of the data points in the SSA component. 

\subsection{LightGBM-based Classification Model} 
LightGBM \cite{ke2017lightgbm} is an ensemble machine learning model based on the technique of gradient boosting decision trees (GBDTs), which is one of the best performing models in machine learning \cite{friedman2001greedy}. LightGBM combines the prediction results of multiple decision trees by iteratively adding them together as follows: 

\begin{equation} \label{GBDT}
    \begin{split}
    \hat{y}_i^{0} &= 0\\
    \hat{y}_i^{1} &= f_1(x_i)=\hat{y}_i^{0}+f_1(x_i)\\
    ...\\
    \hat{y}_i^{s} &= \sum_{k=1}^s f_k(x_i)=\hat{y}_i^{s-1}+f_s(x_i),
    \end{split}
\end{equation}
where $\hat{y}_i^{s}$ is the predicted result for the $i$-th instance at $s$-th iteration and $f_s$ is the learned model for the $s$-th decision tree. This process aimed to minimize a loss function with a regularization term, i.e., 
\begin{equation} \label{GBDT_Obj}
L^{s}(\theta) = \sum_{i}^{n}l(y_i,\hat{y}_i^{s})+\sum_{s=1}^{S}\Omega(f_s),
\end{equation}
where the loss function was binary logloss in this paper and the regularization term controlled the complexity of the model to reduce overfitting. During the training process, GBDTs need to recursively split the data by scanning information gain of each data instance, which is slow when both the sample size and number of the features are big. LightGBM uses a leaf-wise strategy to split the data that has the most information gain using gradient-based one-side sampling (GOSS). The GOSS strategy keeps the samples that contribute the most to information gain with larger gradients and randomly drop those with small gradients. This notion both improves the accuracy and efficiency. LightGBM also uses exclusive feature bundling by combining sparse features that are almost exclusive to each other to further reduce computational resources. These two features made the LightGBM successful and had better performance than XGBoost \cite{chen2016xgboost} across a range of publicly available datasets \cite{ke2017lightgbm}. In this paper, we used the following parameters involved in the training and prediction process of LightGBM: objective: binary, boosting\_type: goss, metric: binary\_logloss, early\_stopping\_rounds: 30, and num\_boost\_round: 500 and used random search for 300 iterations to pick the best combinations among the following parameters, including learning\_rate between 0.01 and 0.5, sub\_feature between 0 and 1, num\_leaves between 5 and 20, min\_data between 10 and 100, max\_depth between 5 and 20. Likewise, we used the following parameters involved in XGBoost as a comparison, including objective: binary:logistic, metric: logloss, early\_stopping\_rounds: 30, and num\_boost\_round: 500 and used random search for 300 iterations to pick the best combinations among the following parameters, including learning rate between 0.01 and 0.5, subsample between 0 and 1, colsample\_bytree between 0 and 1, max\_depth between 5 and 20, and min\_child\_weight between 1 and 10.

\begin{table*}[]
\centering
\caption{Performance comparison between XGBoost, LightGBM, and other methods. }
\label{tab:performance}
\begin{tabular}{ccccccccc}
\hline
Model & Valence & & Arousal &&Liking && EEG & Validation \\ & accuracy    & f1-score    & accuracy    & f1-score    & accuracy    & f1-score    &   &              \\ \hline
XGBoost (all)   & .518 (.016) & .759 (.012) & .542 (.019) & .773 (.021) & .602 (.017) & .825 (.012) & 0 & LOSO         \\
XGBoost (best)  & .570 (.014) & .772 (.012) & .585 (.022) & .785 (.018) & .645 (.020) & .837 (.012) & 0 & LOSO         \\
LightGBM (all)  & .588 (.013) & .804 (.011) & .613 (.022) & .816 (.016) & .667 (.019) & .855 (.011) & 0 & LOSO         \\
\textbf{LightGBM (best)} & .616 (.013) & .814 (.009) & .651 (.018) & .823 (.015) & .691 (.017) & .860 (.011) & 0 & LOSO         \\
SVM \cite{kandemir2014multi}             & .570        & .550        & .560        & .510        & .670        & .450        & 2 & LOSO         \\
MKL \cite{kandemir2014multi}             & .590        & .550        & .560        & .530        & .660        & .510        & 2 & LOSO         \\
MT-MKL (L1) \cite{kandemir2014multi}    & .590        & .550        & .580        & .520        & .660        & .510        & 2 & LOSO         \\
MT-MKL (L2) \cite{kandemir2014multi}    & .600        & .560        & .560        & .460        & .650        & .420        & 2 & LOSO         \\
MESAE \cite{yin2017recognition}     & .830        & .795        & .842        & .780        & -           & -           & 2 & UD (10-fold) \\
CNN \cite{lin2017deep}       & .855        & .801        & .873        & .782        & -           & -           & 2 & UD (10-fold) \\
Bimodal-LSTM \cite{tang2017multimodal}   & .838 (.050) & -           & .832 (.026) & -           & -           & -           & 2 & UD (10-fold) \\
MESAE \cite{yin2017recognition}          & .762        & .724        & .772        & .690        & -           & -           & 2 & 2-fold       \\
Naive Bayes \cite{koelstra2011deap}    & .627        & .605        & .570        & .533        & .591        & .538        & 0 & LOO          \\
Naive Bayes \cite{koelstra2011deap}   & .576        & .563        & .620        & .583        & .554        & .502        & 1 & LOO          \\ \hline
\end{tabular}%
\\
Our results were reported in mean (standard error). "All" indicates when all the features were included and "Best" indicates the best performance when a selected number of the features were included. Other methods were also included for comparisons. However, due to the differences of the experimental settings, it is cautious to directly compare them. Note the column EEG indicates if EEG data from the DEAP dataset were included, where '0' indicated EEG data were not included, '1', indicated only EEG data were included, and '2' indicated both EEG and peripheral physiological data were included. Multiple validation methods were used, where LOSO meant Leave-one-subject-out, UD (10-fold) meant user-dependent (only models from one specific participant were built with) 10-fold cross-validation, and LOO meant leave-one-out cross-validation. MKL = Multiple-Kernel Learning, MT-MKL = Multiple-Task Multiple-Kernel Learning), MESAE = Multiple-fusion-layer based Ensemble classifier of Stacked AutoEncoder.
\end{table*}

\subsection{Model Explanation Using SHAP} 
In order to improve the explainablility of the LightGBM model, we used SHAP \cite{lundberg2017unified,lundberg2018explainable,lundberg2020local}, which was built on Shapley value \cite{shapley1953contributions} from game theory. The Shapley value of a pair of features and its value is computed using:
\begin{equation} \label{shapley}
\varphi_{i}(v)=\sum_{T\subseteq F\setminus \{i\}}{\frac {|T|!(n-|T|-1)!}{n!}}(v(T\cup \{i\})-v(T)), 
\end{equation}
where $n$ is the total number of features, $T$ is a subset of $F$ (i.e., the coalition of all the features), $v(T)$ is the contribution of the coalition $T$ in predicting emotion, and the above equation sums all the subsets $T$ that do not include the feature $i$. Due to the computational load to calculate all the contributions of $T$, when the number of the features is large, SHAP explanation was proposed for tree-based models to reduce its computation from $O(TrL2^n)$ to $O(TrLD^2)$, where $Tr$ is the number of the trees, $L$ is the largest number of the leaves of the trees, $n$ is the total number of the features in the model, and $D$ is the depth of the trees. Due to the fact that SHAP has desirable properties, including local accuracy, missingness, and consistency, which offer a solid theory in explaining machine learning models compared to feature importance measures in other methods \cite{lundberg2017unified}. SHAP is also able to identify the most salient interaction effect between two features, defined as the additional combined feature effects minus individual main feature effects:
\begin{equation} \label{Shap_interaction}
\varphi_{i,j}(v)=\sum_{T\subseteq F\setminus\{i,j\}}\frac{|T|!(n-|T|-2)!}{2(n-1)!}\delta_{ij}(T), 
\end{equation}
where
\begin{equation} \label{coliation}
\delta_{ij}(T)=v(T\cup\{i,j\})-v(T\cup\{i\})-v(T\cup\{j\})+v(T), 
\end{equation}
In this research, SHAP was used to not only identify the most important features in emotion recognition in LightGBM as a feature selection process, but also explain the main effects and interaction effects of the important features. 

\section{Results}
\subsection{Prediction Performance}
We used LOSO cross-validation to show the performance of the proposed method for the three binary classification models, i.e., valence, arousal, and liking. The total sample size was 1280, the training sample size was 1240, and the test sample size was 40 in each round. This process was repeated for 32 times for 32 participants. We used f1-score and accuracy to measure the performance of the models \cite{zhou2011affect}. First, we included all the features in the LightGBM model and SHAP was used to identify the importance of each feature. Second, we added one feature at a time from the most important feature to the least one to find the optimal performance. Table \ref{tab:performance} shows the performance of the proposed model with all the features and an optimal set of features for XGBoost and LightGBM. Using the Wilcoxon signed-rank test, we found that LightGBM performed significantly better than XGBoost for valence, arousal, and liking (all $p<0.001$) both by comparing the results (both accuracy and f1-score) of 51 models (adding one feature at a time, see Fig. \ref{fig:featureSelection}) between LightGBM (best) and XGBoost (best) and between LightGBM (all) and XGBoost (all). The model predicted Liking the best with an average accuracy of 0.691 and an average f1-score of 0.860, followed by arousal and valence with average f1-scores of 0.823 and 0.814, and average accuracy scores of 0.651 and 0.616. As a comparison, we also included the results from another study \cite{kandemir2014multi} that used LOSO cross-validation, showing the potential of the proposed model. 

\subsection{Feature Selection and Importance}
\begin{figure*} [bpt]
	\centering
	\subfloat[\label{fig:valence}]{\includegraphics[width=.32\linewidth]{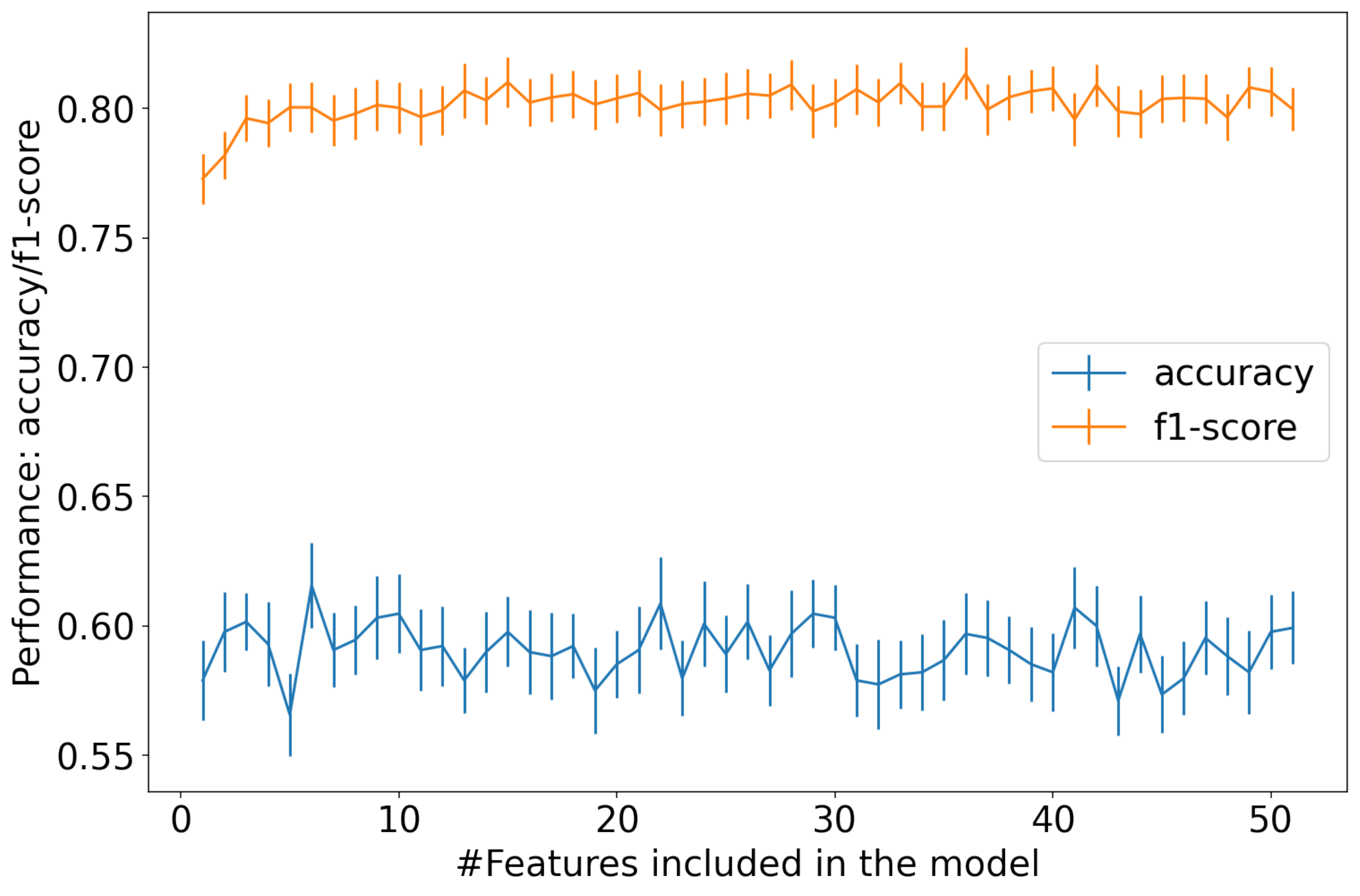}}
	\hspace{0pt}
	\subfloat[\label{fig:arousal}]{\includegraphics[width=.32\linewidth]{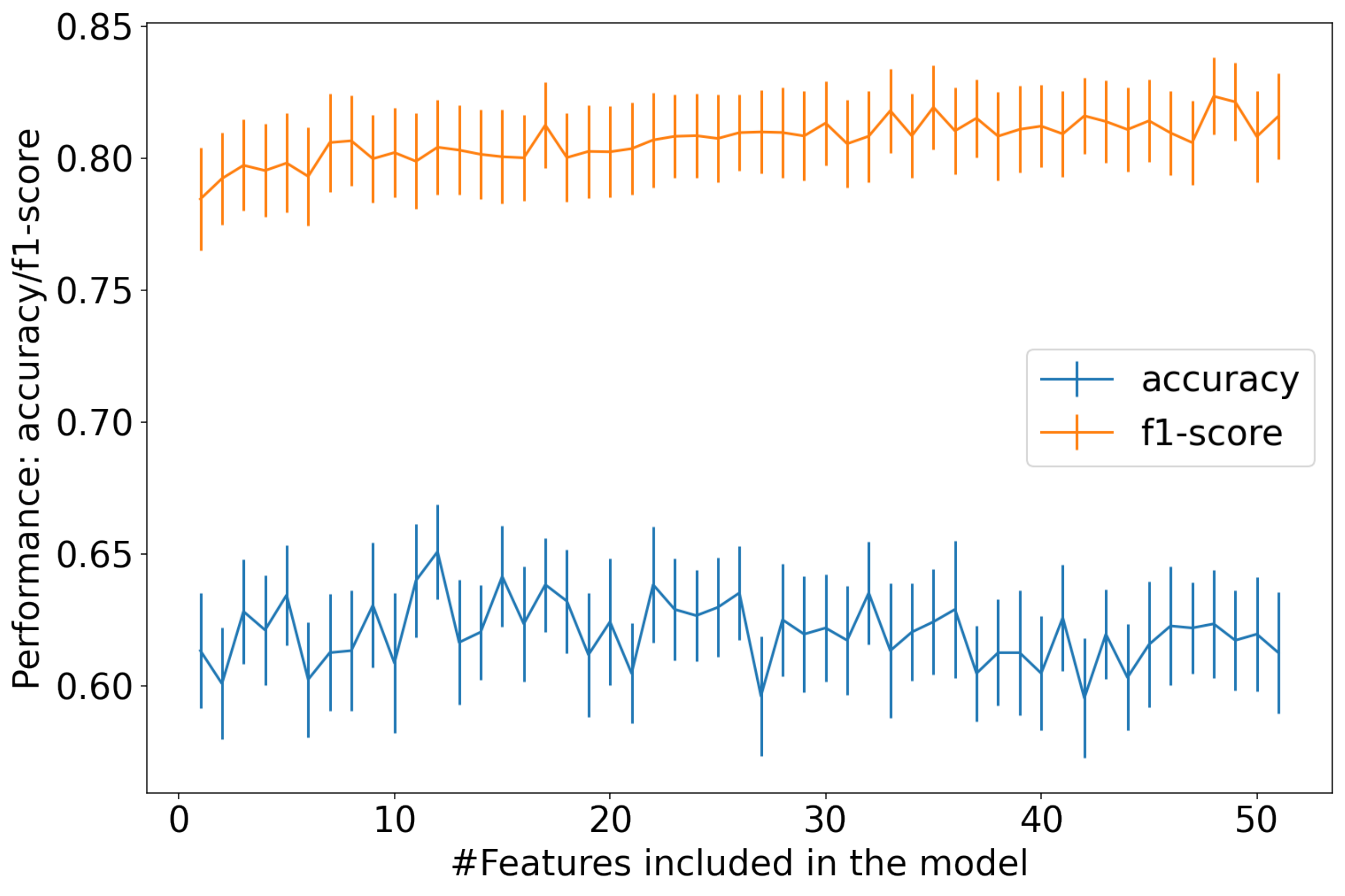}}
	\hspace{0pt}
	\subfloat[\label{fig:liking}]{\includegraphics[width=.32\linewidth]{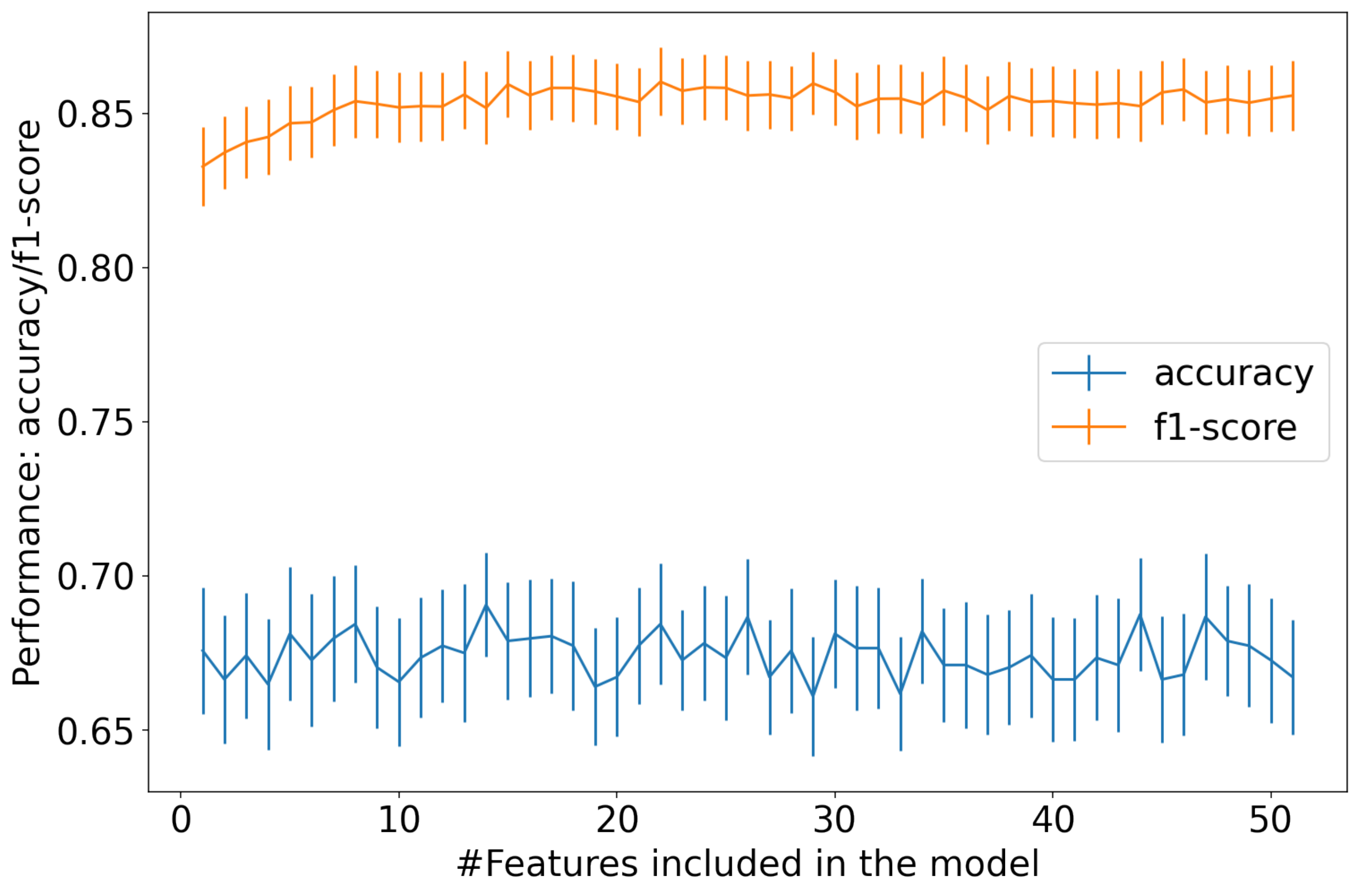}}
    \caption{Feature selection by comparing the performance of the model that added one feature at a time from the most important one to the least important one: (a) Valence; (b) Arousal; (c): Liking. Note the error bar is the standard error produced in the LOSO cross-validation process, the x-axis is the number of the features included in the model and the y-axis is the performance in terms of accuracy and f1-score.}\label{fig:featureSelection}
\end{figure*}

Since LightGBM performed better than XGBoost, the results below were only produced from LightGBM. We first identified the overall ranking of feature importance by including all the features in the LightGBM model. This importance was measured at predicting all the test samples of the 32 rounds in the LOSO process rather than the training samples in order to help improve their generalizability. Then, we obtained the performance of 51 models (corresponding to the total number of the extracted features) by adding one feature at a time from the most important one to the least important one as shown in Fig. \ref{fig:featureSelection}. The best f1-scores were 0.814 (\#feature = 36), 0.823 (\#feature = 48), and 0.860 (\#feature = 22) and the best accuracy was 0.616 (\#feature = 6), 0.651 (\#feature = 12), and 0.691 (\#feature = 14), respectively, for valence, arousal, and liking corresponding to the results in Table \ref{tab:performance}. Therefore, by only selecting a partial number of the features, the best performance was obtained. The variability of accuracy tended to be larger than that of f1-scores for all valence, arousal, and liking. 

\begin{figure*} [hbpt]
	\centering
	\subfloat[\label{fig:valenceFeatures}]{\includegraphics[width=.32\linewidth]{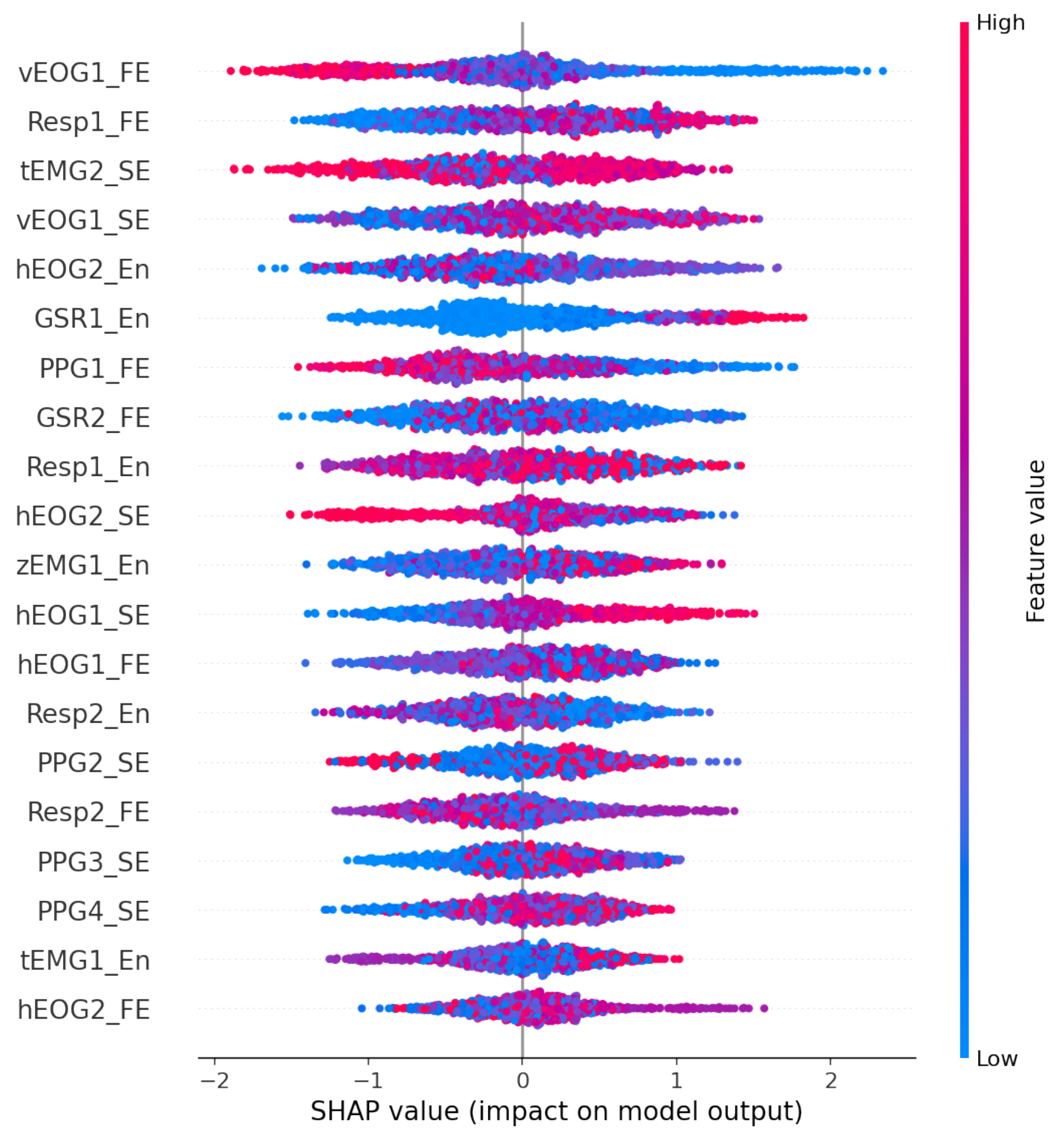}}
	\hspace{0pt}
	\subfloat[\label{fig:arousalFeatures}]{\includegraphics[width=.32\linewidth]{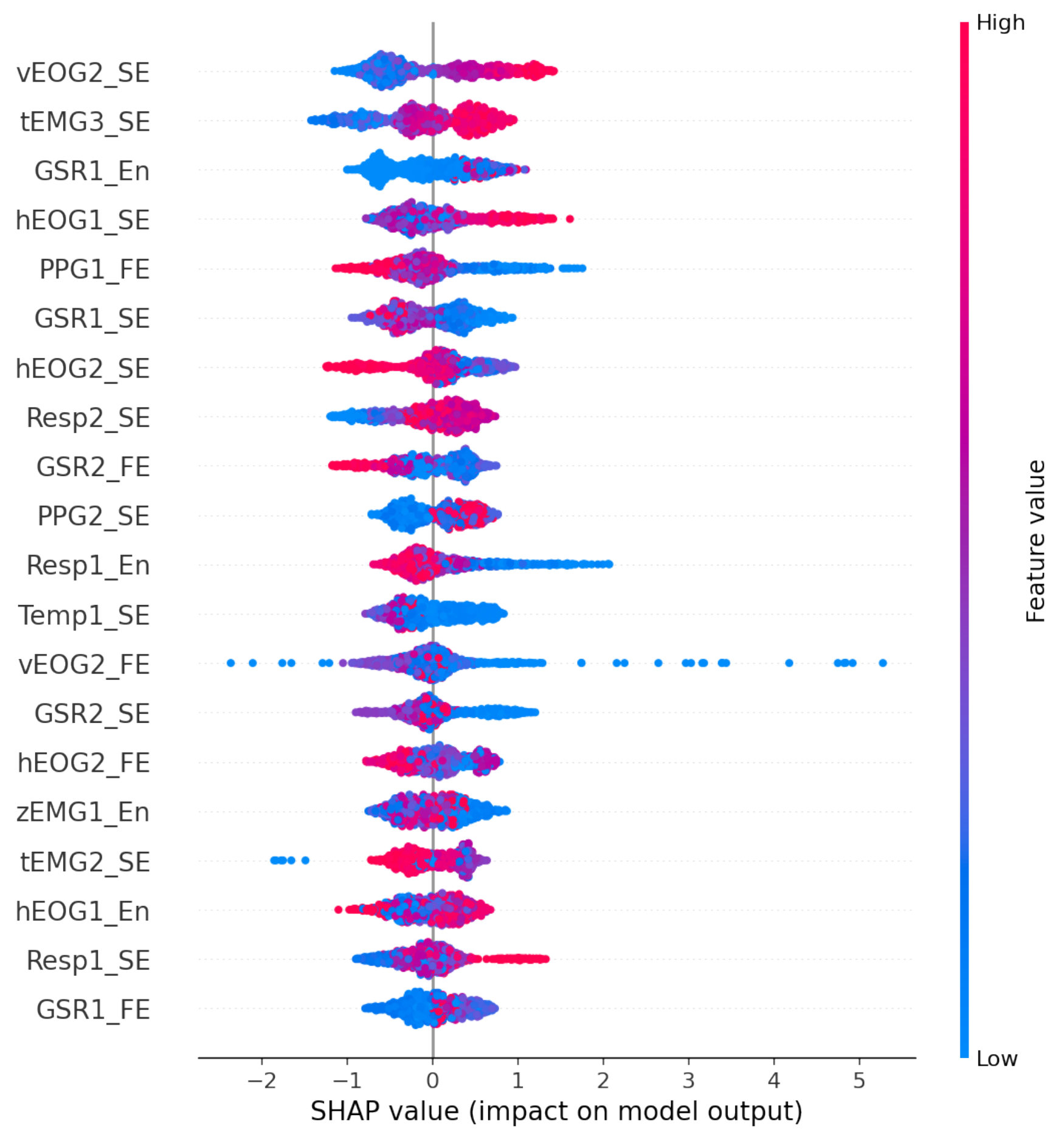}}
	\hspace{0pt}
	\subfloat[\label{fig:likingFeatures}]{\includegraphics[width=.32\linewidth]{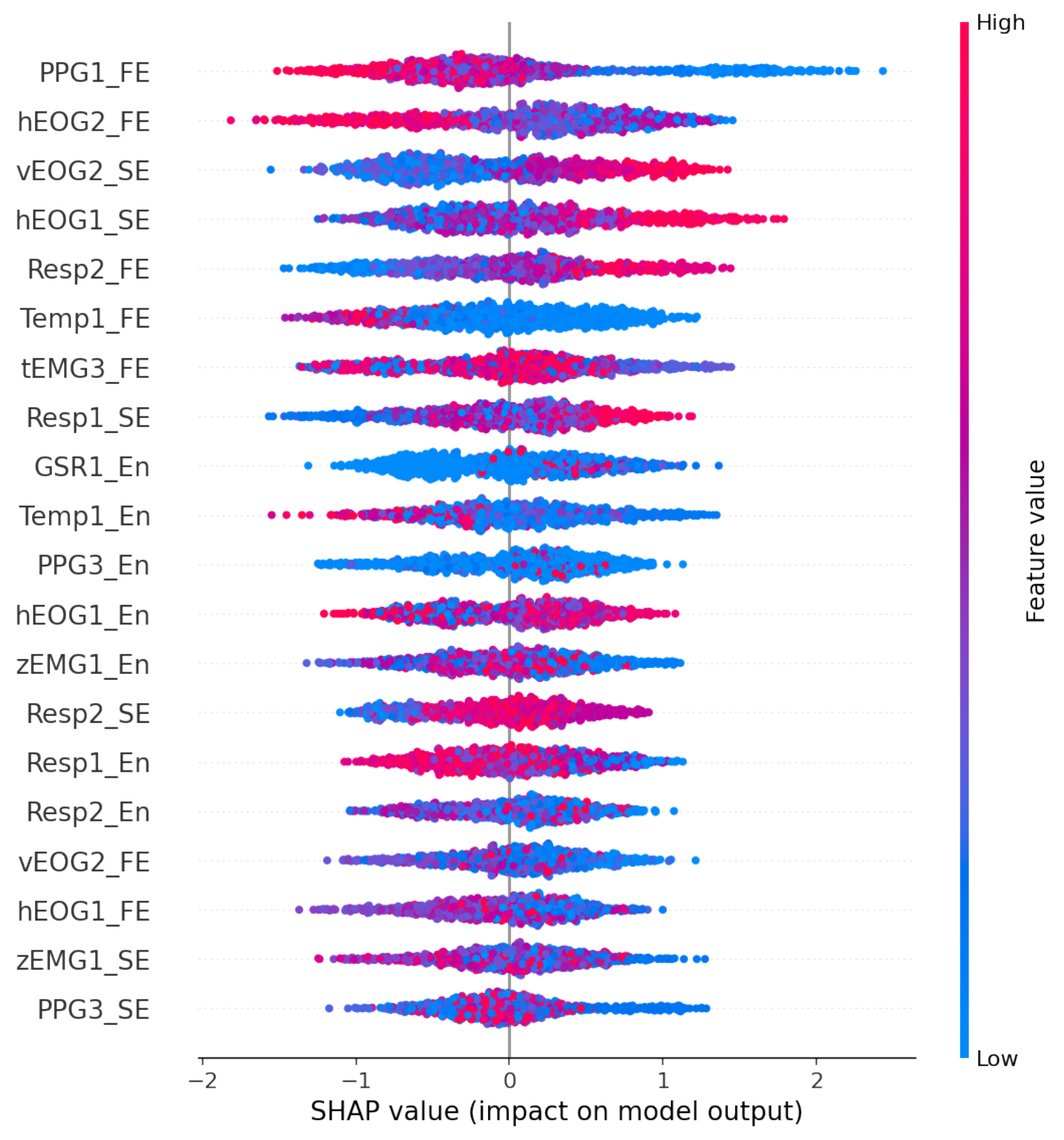}}
    \caption{The top 20 features ranked for their global influence in terms of SHAP values (from top to bottom in descending order) in predicting (a) valence, (b) arousal, and (c) liking using the models with their best f1-scores.  }\label{fig:featureImportance}
\end{figure*}

By using the features selected with the optimal f1-scores, we used SHAP to identify the top 20 features for predicting valence, arousal, and liking in Fig. \ref{fig:featureImportance} ranked by their global influence of each feature in terms of their mean absolute SHAP value (in log loss) using $1/{N_s}\sum_{i=1}^{N_s}|\varphi_i^j|$, where ${N_s}$ is the number of the samples. Each dot (i.e., SHAP value $\varphi_i^j$) is plotted horizontally with color coding. Those in blue are smaller values and those in red are larger values. The horizontal location indicates the contributions (i.e., effects of the variable) to the prediction and the vertical location indicates the global importance of the variable on the prediction. For example, vEOG1\_FE, vEOG2\_SE, and PPG1\_FE were the most important features in predicting valence. The density of the violin figures shows how common the samples are in the dataset. Note in the feature names, the number before "\_" is the corresponding SSA component and the letters after "\_" indicate the feature type, where SE = sample entropy, FE = fuzzy entropy, and En =  Energy. For example, "vEOG1\_FE" indicates the fuzzy entropy of the first component of vertical EOG using SSA decomposition.



\subsection{Main and Interaction Effects}

\begin{figure*} [bpt]
	\centering
	\subfloat[\label{fig:effect_V1}]{\includegraphics[width=0.45\linewidth]{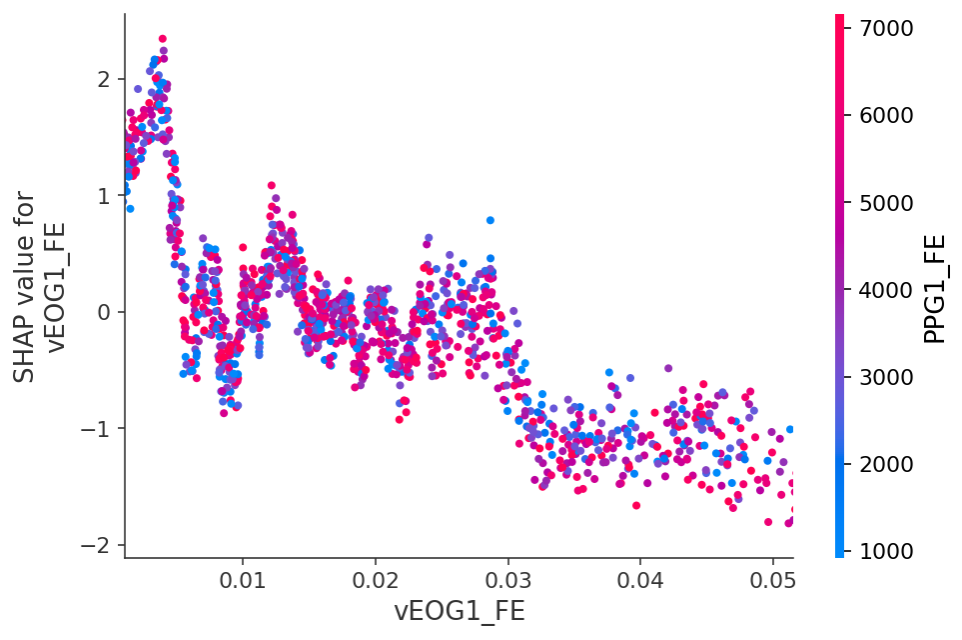}}
	\vspace{0pt}
	\subfloat[\label{fig:effect_V2}]{\includegraphics[width=0.45\linewidth]{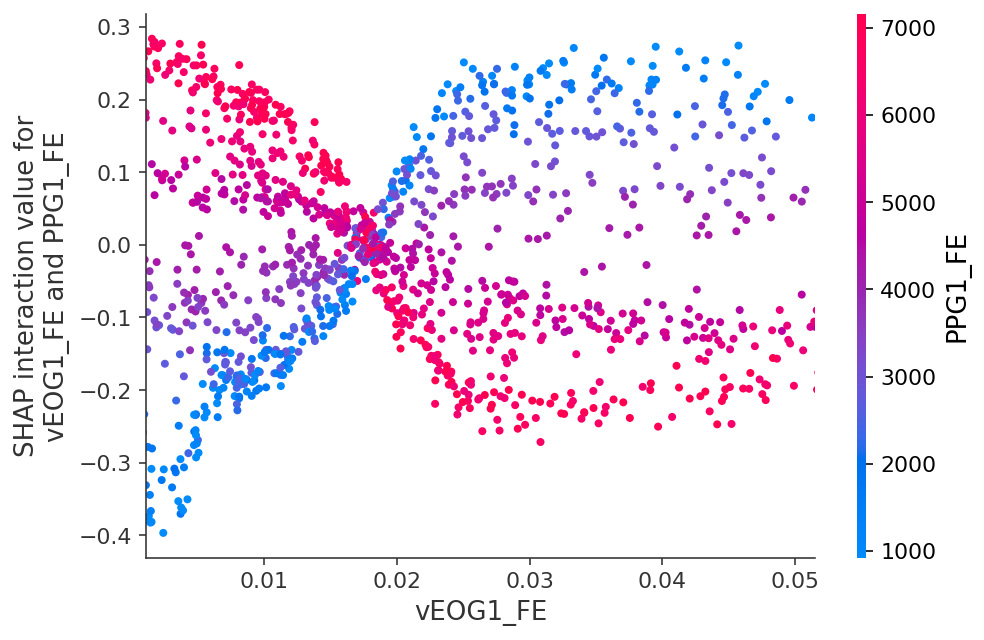}}
    \caption{(a) Main effect of vEOG1\_FE on predicting valence, showing a negative correlation between vEOG1\_FE and contributions to positive valence in terms of its SHAP values. This effect was interacted by another feature named PPG1\_FE and its interaction effect is shown in (b), where larger values of PPG1\_FE leads to more contributions to predicting positive valence than smaller values of PPG1\_FE when vEOG1\_FE is smaller than about 0.018, and this effect is reversed when vEOG1\_FE is larger than about 0.018. Note in both figures the SHAP values for vEOG1\_FE on the left vertical y-axis indicate its contributions to valence and a positive value leads to more likelihood of positive valence; the main effect of vEOG1\_FE is interacted by PPG1\_FE located on the right vertical y-axis and its values are color-coded, i.e., larger ones are in red while smaller ones are in blue. This also applies to Fig. \ref{fig:effectsA} and Fig. \ref{fig:effectsL}. }\label{fig:effectsV}
\end{figure*}
\begin{figure*} [bpt]
	\centering
	\subfloat[\label{fig:effect_A1}]{\includegraphics[width=0.45\linewidth]{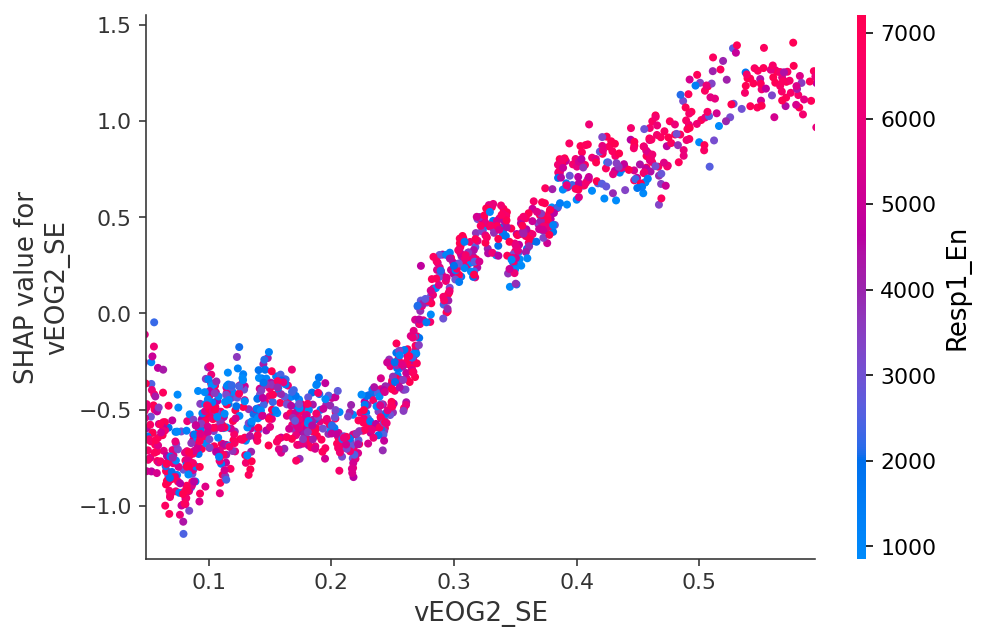}}
	\vspace{0pt}
	\subfloat[\label{fig:effect_A2}]{\includegraphics[width=0.45\linewidth]{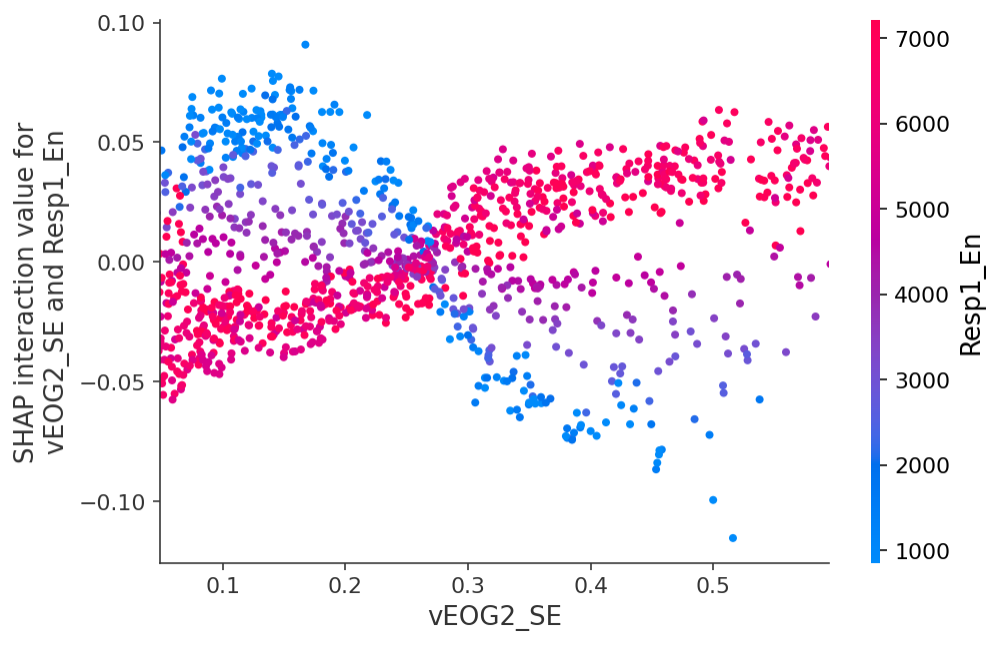}}
    \caption{(a) Main effect of vEOG2\_SE on predicting arousal, showing a positive correlation between vEOG2\_SE and contributions to high arousal in terms of its SHAP values. This effect was interacted by another feature named Resp1\_En, and its interaction effect is shown in (b) where larger values of Resp1\_En leads to more contributions to predicting low arousal  than smaller values of Resp1\_En, when vEOG2\_SE is smaller than about 0.28 and this effect is reversed when vEOG2\_SE is larger than about 0.28.}\label{fig:effectsA}
\end{figure*}
\begin{figure*} [bpt]
	\centering
	\subfloat[\label{fig:effect_L1}]{\includegraphics[width=0.45\linewidth]{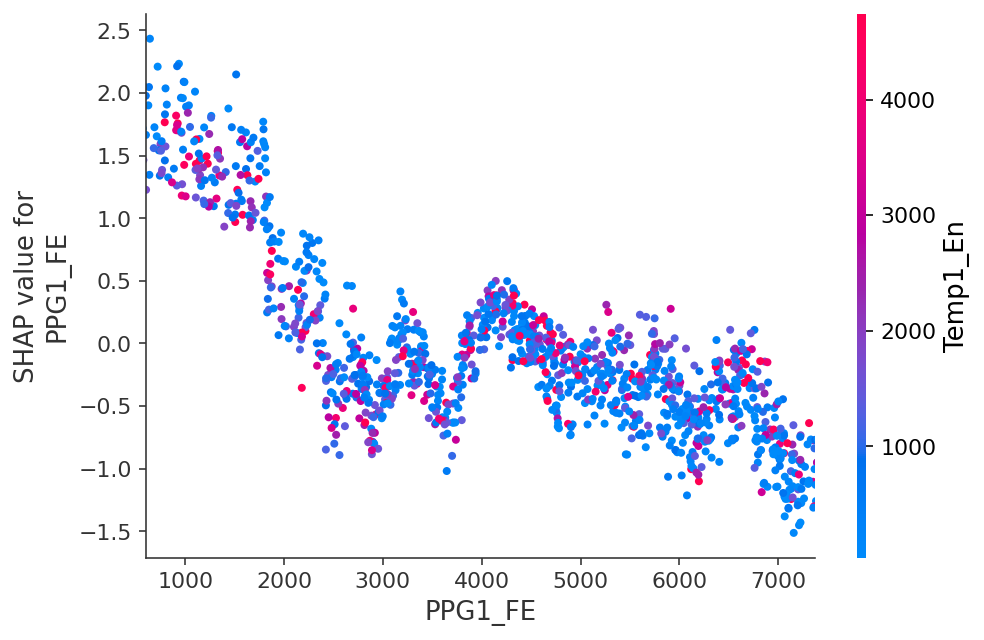}}
	\vspace{0pt}
	\subfloat[\label{fig:effect_L2}]{\includegraphics[width=0.45\linewidth]{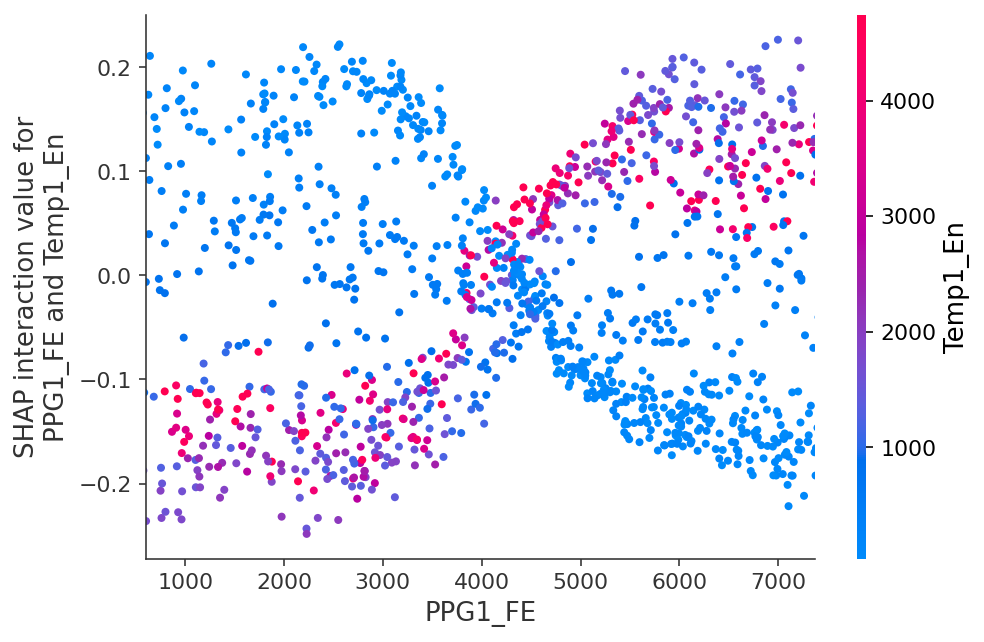}}
    \caption{(a) Main effect of PPG1\_FE on predicting liking, showing a negative correlation between PPG1\_FE and the contributions to liking in terms of its SHAP values. This effect was interacted by another feature named Temp1\_En and its interaction effect is shown in (b) where smaller values of Temp1\_En leads to more contributions to predicting liking  than larger values of Temp1\_En, when PPG1\_FE is smaller than about 4000 and this effect is reversed when PPG1\_FE is larger than about 4000.}\label{fig:effectsL}
\end{figure*}

\begin{figure*} [bpt]
	\centering
	\subfloat[\label{fig:valenceInteraction}]{\includegraphics[width=.32\linewidth]{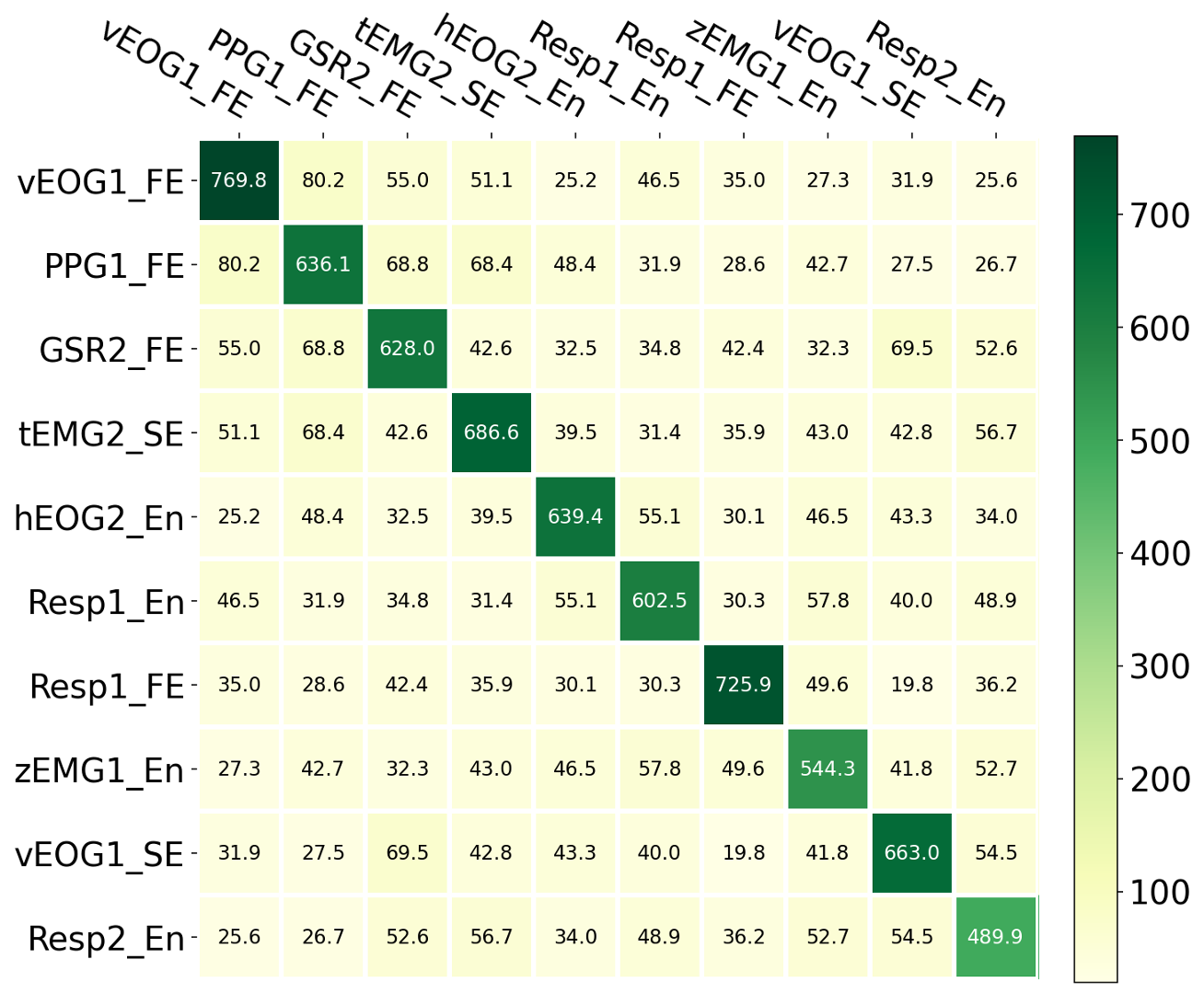}}
	\hspace{0pt}
	\subfloat[\label{fig:arousalInteraction}]{\includegraphics[width=.32\linewidth]{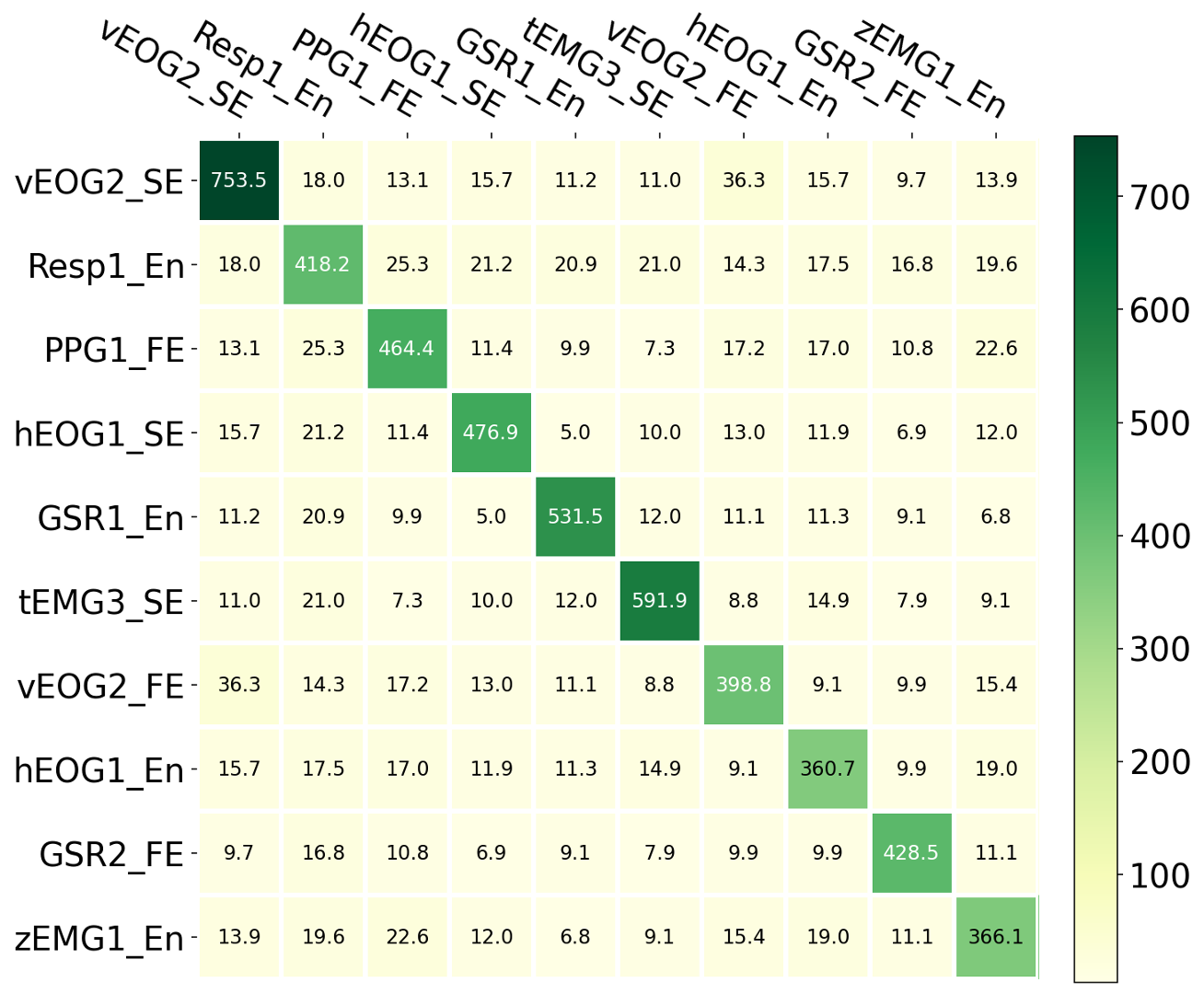}}
	\hspace{0pt}
	\subfloat[\label{fig:likingInteraction}]{\includegraphics[width=.32\linewidth]{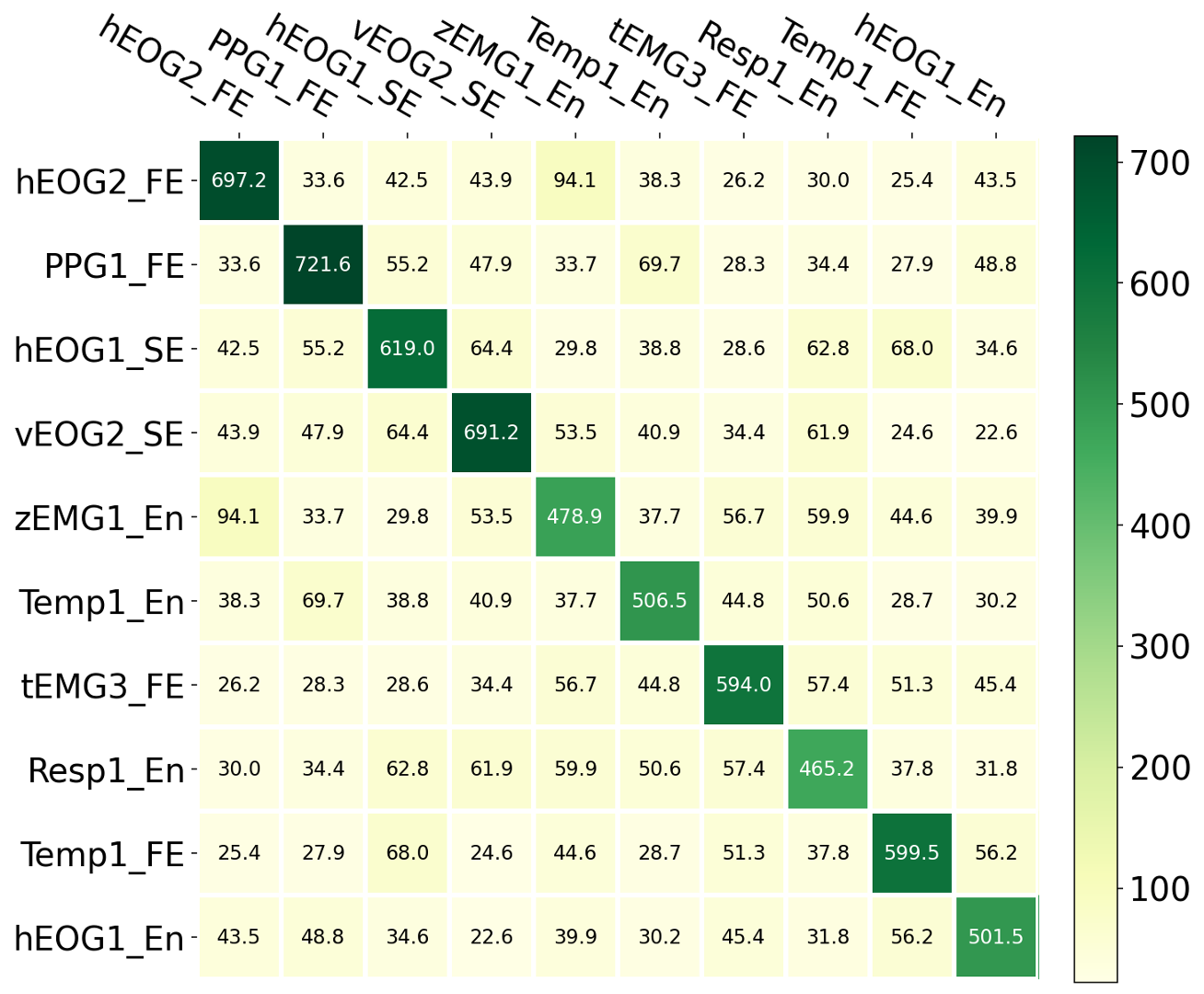}}
    \caption{Interaction matrices for the top 10 features that had the strongest total interaction with other features: (a) Valence interaction matrix, (b) Arousal interaction matrix, (c) Liking interaction matrix.}\label{fig:interactionMatrix}
\end{figure*}

To better understand how individual features influence the model prediction, we further examined the main and interaction effects. Due to space limitations, we only showed one main and interaction effect each for valence, arousal, and liking. Note we only included data from the 2.5th percentile to 97.5th percentile in order to examine the major trend. Fig. \ref{fig:effect_V1} shows the main effect of vEOG1\_FE. The overall trend is that the larger the value of vEOG1\_FE, the less likely the positive valence. This is consistent with the results in Fig. \ref{fig:valenceFeatures}. Note a larger entropy value in this paper indicates the signal is less regular, and vice versa. At the same time, vEOG1\_FE interacts with PPG1\_FE. In order to better observe the interaction effect, we show it in Fig. \ref{fig:effect_V2} that larger values of PPG1\_FE leads to more contributions to predicting positive valence than smaller values of PPG1\_FE when vEOG1\_FE is smaller than about 0.018, and this effect is reversed when vEOG1\_FE is larger than about 0.018. This interaction effect was also calculated as 80.2 as shown in Fig. \ref{fig:valenceInteraction} using Eq. \ref{Shap_interaction}. 
Fig. \ref{fig:effect_A1} shows the main effect of vEOG2\_SE on predicting arousal, showing a positive correlation between vEOG2\_SE and contributions to high arousal in terms of its SHAP values. This effect was interacted by another feature named Resp1\_En and its interaction effect is shown in Fig. \ref{fig:effect_A2} where larger values of Resp1\_En leads to more contributions to predicting low arousal than smaller values of Resp1\_En, when vEOG2\_SE is smaller than about 0.28 and this effect is reversed when vEOG2\_SE is larger than about 0.28. This interaction effect was also calculated as 18.0 as shown in Fig. \ref{fig:arousalInteraction} using Eq. \ref{Shap_interaction}. 
Fig. \ref{fig:effect_L1} shows the main effect of PPG1\_FE on predicting liking, showing a negative correlation between PPG1\_FE and the contributions to liking in terms of its SHAP values. This effect was interacted by another feature named Temp1\_En and its interaction effect is shown in Fig. \ref{fig:effect_L2} where smaller values of Temp1\_En leads to more contributions to predicting liking  than larger values of Temp1\_En, when PPG1\_FE is smaller than about 4000 and this effect is reversed when PPG1\_FE is larger than about 4000. This interaction effect was also calculated as 69.7 as shown in Fig. \ref{fig:likingInteraction} using Eq. \ref{Shap_interaction}. 

In order to show the interaction effects between different features of the LightGBM (best) model as shown in Table \ref{tab:performance}, we calculated the interaction matrices for valence, arousal, and liking in Fig. \ref{fig:interactionMatrix} ordered by features of total interaction effects with all other features selected in the model. Those in the diagonal entries indicate the main effects of the specific features, while those in the off-diagonal entries indicate the interaction effects between two features. Note only features of the total interaction effects in the top ten list were shown in the matrix due to space limit. As an example, the top three features are  vEOG1\_FE, PPG1\_FE, and GSR2\_FE in Fig. \ref{fig:valenceInteraction}, and the feature vEOG1\_FE has the largest total interaction effects by summing all other interaction effects with others. 

\section{Discussion}
\subsection{Model Performance}
In this paper, we used a LOSO cross-validation process in order to improve the generalizability of the explanations of the LightGBM models. Compared to previous studies \cite{kandemir2014multi,romeo2019multiple} with LOSO cross-validation, our proposed method had better performance. Despite the fact that there were better results reported using the same dataset, they did not adopt a LOSO cross-validation process or also used EEG data (see Table \ref{tab:performance}). The application of SSA was effective to improve the performance of our models, which was reported to eliminate noise in the raw data better than digital filtering methods \cite{alonso2005application}, and decomposed the raw data into independent additive components \cite{elsner2013singular}.  By making use of the  singular value hard thresholding formula \cite{6846297}, we were able to remove the noise effectively. The SSA method was also shown to improve the classification performance of machine learning models constructed from physiological signals \cite{liu2008physiology,lu2020dynamic}. In addition, there were only two parameters involved in SSA decomposition, including the window length and the number of the components for reconstruction, which made it easier for practical uses. Another important reason that improves the performance of our models might be the extracted features. Rather than using a statistic features (e.g., mean, standard deviation), we made use of complexity and energy measures. For one thing, entropy measures the randomness of a time series signal, which is an essential limitation of moment statistics \cite{delgado2019approximate}. For example, a periodic signal (e.g., RC 2 in Fig. \ref{fig:PPG2}) could have exactly the same mean value and standard deviation if it was randomized while the entropy measures are able to tell the differences between the two. Similarly, there were only a small number of parameters in calculating entropy measures, including the embedding dimension and the tolerance for sample entropy, and a third parameter, i.e., the step of the fuzzy function. We selected these parameters based on previous studies \cite{chen2007characterization,lu2020dynamic,lu2020entropy}, which proved to have good performance in this paper. Moreover, the inclusion of energy feature was also helpful, as evidenced in Fig. \ref{fig:featureImportance}, where there are 6, 4, and 7 energy features (those ended with "\_En") included in the top 20 features for valence, arousal, and liking, respectively. Furthermore, LightGBM excelled in emotion recognition using the extracted features in terms of effectiveness and efficiency. The GOSS strategy and the feature bundle approach played a significant role in improving the performance compared to another successful tree-ensemble method, i.e., XGBoost. Finally, the feature selection method enabled by SHAP boosted the performance of emotion recognition as evidenced in Fig. \ref{fig:featureSelection}. Hence, the good performance was the result by leveraging different aspects of the proposed method.

\subsection{Explainability for Emotion Recognition}
Our research is among the few examples that attempted to explain the hidden knowledge captured by the machine learning models in emotion recognition. First, SHAP was able to identify the most important features. As shown in Fig. \ref{fig:featureImportance}, the top three most important features are vEOG1\_FE, Resp1\_FE and tEMG2\_SE for valence prediction, vEOG2\_SE, tEMG3\_SE, and GSR1\_En for arousal prediction, and PPG1\_FE, hEGO2\_FE, and vEOG2\_SE for liking prediction. Also as shown in Fig. \ref{fig:featureSelection}, a small number of top features already have good performance in terms of f1-scores. Such insights are extremely helpful in selecting features and improving model performance. In addition, the numbers of SSA components decomposed from each raw signal were small. There were two, two, one, three, two, two, one, and four components extracted from hEOG, vEOG, zEMG, tEMG, SCR, Resp, and PPG. Such SSA components were either varying trends or oscillatory signals (see Fig. \ref{fig:SSA}). This helped interpret individual features to a large extent. For example, vEOG1\_FE  and vEOG2\_FE were the fuzzy entropy of the major varying trend and the periodic component of vEOG and the trend component was most important in predicting valence in the model. Even though there were four oscillatory components extracted from PPG signals, they might be related to cardiac changes in the blood volume related to heart beats, superimposed by other components due to respiration, thermo-regulation, and activities in the sympathetic nervous system \cite{SUBASI201927}. Further investigation is needed to identify their mapping relations.
Second, we are able to identify the hidden relationships between different features and emotions (i.e., valence, arousal, and liking). For example, vEOG1\_FE was negatively correlated with its contributions (i.e., SHAP values) to positive valence prediction in Fig. \ref{fig:effect_V1}, vEOG2\_SE was positively correlated with its contributions to high arousal prediction in Fig. \ref{fig:effect_A1}, and PPG1\_FE was negatively correlated with its contributions to liking prediction in Fig. \ref{fig:effect_L1}. By examining such relationships, we are able to understand the unique patterns of physiological measures associated with different emotions and such insights are useful for different practical applications as evidenced by \cite{weitz2019deep}. Thus, the most important features derived from physiological data could recognize emotion reliably and consistently, based on which appropriate measures might be provided in various areas for effective emotion intervention, such as medical care \cite{doukas2008intelligent}. Third, what is less well known is the interaction between different features that SHAP identified as shown in Figs. \ref{fig:effect_V2}, \ref{fig:effect_A2}, and \ref{fig:effect_L2}. Such interaction effects were visualized continuously rather than among a small number of levels in classical statistical analysis. We further calculated the interaction matrices in Fig. \ref{fig:interactionMatrix} and we quantitatively calculated the interaction between each features, which offered a big picture in understanding how individual features influence emotion classification.  

\subsection{Limitations and Future Work}
Our study also suffers from limitations, which can be left for future work. First, the SSA method helped extract useful features to improve model performance. However, one feature was decomposed into several and the relationships between these decomposed components and emotions can be less clear compared to the original raw signal. For example, we found that PPG2\_SE, PPG3\_SE, and PPG4\_SE were found to be in the top 20 features for valence prediction. As mentioned previously, more investigation is needed to understand what these periodic components mean physiologically as shown in Fig. \ref{fig:valenceFeatures}. Furthermore, it might be less clear about the overall relationship between PPG and valence as a whole without examining all the four components of PPG. One possible solution is to further reduce the number of the features in the model. As shown in Fig. \ref{fig:featureSelection}, the performance of models can be quickly stabilized with only a small number of features in terms of f1-scores.  Second, the features were extracted from 60-second time series samples, which enabled the LightGMB model to perform the classification purposes. However, we were not able to tell the dynamics of the physiological data. Future research can be conducted by using a sliding time window, which can potentially improve the performance \cite{lu2020dynamic}. Third, we selected the values of the parameters involved in SSA and entropy measures based on previous studies \cite{chen2007characterization,lu2020dynamic,lu2020entropy}. 
In the future, it is possible to conduct a sensitivity analysis with regard to different parameters involved, including the window length and number of components in SSA and the embedding dimension in calculating entropy measures, to help understand how they influence the model performance. Fourth, we used the DEAP dataset to demonstrate the proposed method. Specifically, we only used the 12 peripheral physiological data for emotion recognition with LOSO cross validation. In the future, EEG signals can also be included to check the performance of the proposed method. The dataset was recorded in two separate locations, including Twente and Geneva. It potentially can help improve the capability of generalizing the results, but also had differences in data quality due to different hardware. Our proposed method can also be applied to other datasets to examine its generalizability.

\section{Conclusion}
In this paper, we proposed a method to recognize emotions using peripheral physiological data by taking both performance and explainability into consideration. First, in order to deal with the spare labeling issue in the physiological data, we used both complexity and energy measures to extract features from decomposed components enabled by SSA. Such an approach proved to be effective in this paper as evidenced by the better performance of LightGBM when it was compared with XGBoost and previous research. Second, in order to improve the explainability of the LightGBM model, SHAP was used to identify the most important features and the relationships (i.e., main and interaction effects) between individual features and emotions. 
Therefore, our proposed method has good implications on not only improving the performance of emotion recognition, but also on deploying the model for affective human-machine systems.


%

\ifCLASSOPTIONcaptionsoff
  \newpage
\fi



%
\bibliographystyle{IEEEtran}
\bibliography{IEEEabrv,reference}

\begin{thebibliography}{10}
\providecommand{\url}[1]{#1}
\csname url@samestyle\endcsname
\providecommand{\newblock}{\relax}
\providecommand{\bibinfo}[2]{#2}
\providecommand{\BIBentrySTDinterwordspacing}{\spaceskip=0pt\relax}
\providecommand{\BIBentryALTinterwordstretchfactor}{4}
\providecommand{\BIBentryALTinterwordspacing}{\spaceskip=\fontdimen2\font plus
\BIBentryALTinterwordstretchfactor\fontdimen3\font minus
  \fontdimen4\font\relax}
\providecommand{\BIBforeignlanguage}[2]{{%
\expandafter\ifx\csname l@#1\endcsname\relax
\typeout{** WARNING: IEEEtran.bst: No hyphenation pattern has been}%
\typeout{** loaded for the language `#1'. Using the pattern for}%
\typeout{** the default language instead.}%
\else
\language=\csname l@#1\endcsname
\fi
#2}}
\providecommand{\BIBdecl}{\relax}
\BIBdecl

\bibitem{picard2000affective}
R.~W. Picard, \emph{Affective computing}.\hskip 1em plus 0.5em minus
  0.4em\relax MIT press, 2000.

\bibitem{zhou2014prospect}
F.~Zhou, Y.~Ji, and R.~J. Jiao, ``Prospect-theoretic modeling of customer
  affective-cognitive decisions under uncertainty for user experience design,''
  \emph{IEEE Transactions on Human-Machine Systems}, vol.~44, no.~4, pp.
  468--483, 2014.

\bibitem{yerkes1908relation}
R.~M. Yerkes, J.~D. Dodson \emph{et~al.}, ``The relation of strength of
  stimulus to rapidity of habit-formation,'' \emph{Punishment: Issues and
  experiments}, pp. 27--41, 1908.

\bibitem{du2020examining}
N.~Du, F.~Zhou, E.~M. Pulver, D.~M. Tilbury, L.~P. Robert, A.~K. Pradhan, and
  X.~J. Yang, ``Examining the effects of emotional valence and arousal on
  takeover performance in conditionally automated driving,''
  \emph{Transportation research part C: emerging technologies}, vol. 112, pp.
  78--87, 2020.

\bibitem{chen2017dynamic}
L.~Chen, M.~Wu, M.~Zhou, Z.~Liu, J.~She, and K.~Hirota, ``Dynamic emotion
  understanding in human--robot interaction based on two-layer fuzzy svr-ts
  model,'' \emph{IEEE Transactions on Systems, Man, and Cybernetics: Systems},
  vol.~50, no.~2, pp. 490--501, 2017.

\bibitem{zhou2011affect}
F.~Zhou, X.~Qu, M.~G. Helander, and J.~R. Jiao, ``Affect prediction from
  physiological measures via visual stimuli,'' \emph{International Journal of
  Human-Computer Studies}, vol.~69, no.~12, pp. 801--819, 2011.

\bibitem{zhou2020fine}
F.~Zhou, S.~Kong, C.~C. Fowlkes, T.~Chen, and B.~Lei, ``Fine-grained facial
  expression analysis using dimensional emotion model,'' \emph{Neurocomputing},
  vol. 397, pp. 38--49, 2020.

\bibitem{sun2019design}
X.~Sun, Z.~Pei, C.~Zhang, G.~Li, and J.~Tao, ``Design and analysis of a
  human-machine interaction system for researching human's dynamic emotion,''
  \emph{IEEE Transactions on Systems, Man, and Cybernetics: Systems}, 2019.

\bibitem{zhou2014emotion}
F.~Zhou, X.~Qu, J.~Jiao, and M.~G. Helander, ``Emotion prediction from
  physiological signals: A comparison study between visual and auditory
  elicitors,'' \emph{Interacting with computers}, vol.~26, no.~3, pp. 285--302,
  2014.

\bibitem{zhou2020driver}
F.~Zhou, A.~Alsaid, M.~Blommer, R.~Curry, R.~Swaminathan, D.~Kochhar,
  W.~Talamonti, L.~Tijerina, and B.~Lei, ``Driver fatigue transition prediction
  in highly automated driving using physiological features,'' \emph{Expert
  Systems with Applications}, p. 113204, 2020.

\bibitem{stone1994ecological}
A.~A. Stone and S.~Shiffman, ``Ecological momentary assessment (ema) in
  behavorial medicine.'' \emph{Annals of Behavioral Medicine}, 1994.

\bibitem{zhao2019speech}
J.~Zhao, X.~Mao, and L.~Chen, ``Speech emotion recognition using deep 1d \& 2d
  cnn lstm networks,'' \emph{Biomedical Signal Processing and Control},
  vol.~47, pp. 312--323, 2019.

\bibitem{frantzidis2010classification}
C.~A. Frantzidis, C.~Bratsas, M.~A. Klados, E.~Konstantinidis, C.~D. Lithari,
  A.~B. Vivas, C.~L. Papadelis, E.~Kaldoudi, C.~Pappas, and P.~D. Bamidis, ``On
  the classification of emotional biosignals evoked while viewing affective
  pictures: an integrated data-mining-based approach for healthcare
  applications,'' \emph{IEEE Transactions on Information Technology in
  Biomedicine}, vol.~14, no.~2, pp. 309--318, 2010.

\bibitem{schiano2004categorical}
D.~J. Schiano, S.~M. Ehrlich, and K.~Sheridan, ``Categorical imperative not:
  facial affect is perceived continuously,'' in \emph{Proceedings of the SIGCHI
  conference on Human factors in computing systems}, 2004, pp. 49--56.

\bibitem{zhou2013affective}
F.~Zhou, Y.~Ji, and R.~J. Jiao, ``Affective and cognitive design for mass
  personalization: status and prospect,'' \emph{Journal of Intelligent
  Manufacturing}, vol.~24, no.~5, pp. 1047--1069, 2013.

\bibitem{zhou2020emotional}
------, ``Emotional design,'' \emph{arXiv preprint arXiv:2010.03046}, 2020.

\bibitem{kim2008emotion}
J.~Kim and E.~Andr{\'e}, ``Emotion recognition based on physiological changes
  in music listening,'' \emph{IEEE transactions on pattern analysis and machine
  intelligence}, vol.~30, no.~12, pp. 2067--2083, 2008.

\bibitem{walter2013transsituational}
S.~Walter, J.~Kim, D.~Hrabal, S.~C. Crawcour, H.~Kessler, and H.~C. Traue,
  ``Transsituational individual-specific biopsychological classification of
  emotions,'' \emph{IEEE Transactions on Systems, Man, and Cybernetics:
  Systems}, vol.~43, no.~4, pp. 988--995, 2013.

\bibitem{6846297}
M.~{Gavish} and D.~L. {Donoho}, ``The optimal hard threshold for singular
  values is $4/\sqrt {3}$,'' \emph{IEEE Transactions on Information Theory},
  vol.~60, no.~8, pp. 5040--5053, 2014.

\bibitem{wen2014emotion}
W.~Wen, G.~Liu, N.~Cheng, J.~Wei, P.~Shangguan, and W.~Huang, ``Emotion
  recognition based on multi-variant correlation of physiological signals,''
  \emph{IEEE Transactions on Affective Computing}, vol.~5, no.~2, pp. 126--140,
  2014.

\bibitem{ma2019emotion}
J.~Ma, H.~Tang, W.-L. Zheng, and B.-L. Lu, ``Emotion recognition using
  multimodal residual lstm network,'' in \emph{Proceedings of the 27th ACM
  International Conference on Multimedia}, 2019, pp. 176--183.

\bibitem{romeo2019multiple}
L.~Romeo, A.~Cavallo, L.~Pepa, N.~Berthouze, and M.~Pontil, ``Multiple instance
  learning for emotion recognition using physiological signals,'' \emph{IEEE
  Transactions on Affective Computing}, vol.~13, no.~01, pp. 1--1, nov 2019.

\bibitem{lan2019domain}
Z.~Lan, O.~Sourina, L.~Wang, R.~Scherer, and G.~R. Müller-Putz, ``Domain
  adaptation techniques for eeg-based emotion recognition: A comparative study
  on two public datasets,'' \emph{IEEE Transactions on Cognitive and
  Developmental Systems}, vol.~11, no.~1, pp. 85--94, 2019.

\bibitem{li2020multisource}
J.~Li, S.~Qiu, Y.-Y. Shen, C.-L. Liu, and H.~He, ``Multisource transfer
  learning for cross-subject eeg emotion recognition,'' \emph{IEEE Transactions
  on Cybernetics}, vol.~50, no.~7, pp. 3281--3293, 2020.

\bibitem{chen2021personal}
H.~Chen, S.~Sun, J.~Li, R.~Yu, N.~Li, X.~Li, and B.~Hu, ``Personal-zscore:
  Eliminating individual difference for eeg-based cross-subject emotion
  recognition,'' \emph{IEEE Transactions on Affective Computing}, pp. 1--1,
  2021.

\bibitem{koelstra2011deap}
S.~Koelstra, C.~Muhl, M.~Soleymani, J.-S. Lee, A.~Yazdani, T.~Ebrahimi, T.~Pun,
  A.~Nijholt, and I.~Patras, ``Deap: A database for emotion analysis; using
  physiological signals,'' \emph{IEEE transactions on affective computing},
  vol.~3, no.~1, pp. 18--31, 2011.

\bibitem{soleymani2011multimodal}
M.~Soleymani, J.~Lichtenauer, T.~Pun, and M.~Pantic, ``A multimodal database
  for affect recognition and implicit tagging,'' \emph{IEEE transactions on
  affective computing}, vol.~3, no.~1, pp. 42--55, 2011.

\bibitem{zheng2018emotionmeter}
W.-L. Zheng, W.~Liu, Y.~Lu, B.-L. Lu, and A.~Cichocki, ``Emotionmeter: A
  multimodal framework for recognizing human emotions,'' \emph{IEEE
  transactions on cybernetics}, vol.~49, no.~3, pp. 1110--1122, 2018.

\bibitem{mollahosseini2017affectnet}
A.~Mollahosseini, B.~Hasani, and M.~H. Mahoor, ``Affectnet: A database for
  facial expression, valence, and arousal computing in the wild,'' \emph{IEEE
  Transactions on Affective Computing}, vol.~10, no.~1, pp. 18--31, 2017.

\bibitem{yannakakis2008entertainment}
G.~N. Yannakakis and J.~Hallam, ``Entertainment modeling through physiology in
  physical play,'' \emph{International Journal of Human-Computer Studies},
  vol.~66, no.~10, pp. 741--755, 2008.

\bibitem{li2017automatic}
Z.~Li, L.~Chen, J.~Peng, and Y.~Wu, ``Automatic detection of driver fatigue
  using driving operation information for transportation safety,''
  \emph{Sensors}, vol.~17, no.~6, p. 1212, 2017.

\bibitem{doshi2017towards}
F.~Doshi-Velez and B.~Kim, ``Towards a rigorous science of interpretable
  machine learning,'' \emph{arXiv preprint arXiv:1702.08608}, 2017.

\bibitem{lundberg2018explainable}
S.~M. Lundberg, B.~Nair, M.~S. Vavilala, M.~Horibe, M.~J. Eisses, T.~Adams,
  D.~E. Liston, D.~K.-W. Low, S.-F. Newman, J.~Kim \emph{et~al.}, ``Explainable
  machine-learning predictions for the prevention of hypoxaemia during
  surgery,'' \emph{Nature biomedical engineering}, vol.~2, no.~10, pp.
  749--760, 2018.

\bibitem{yan2020interpretable}
L.~Yan, H.-T. Zhang, J.~Goncalves, Y.~Xiao, M.~Wang, Y.~Guo, C.~Sun, X.~Tang,
  L.~Jing, M.~Zhang \emph{et~al.}, ``An interpretable mortality prediction
  model for covid-19 patients,'' \emph{Nature Machine Intelligence}, pp. 1--6,
  2020.

\bibitem{AYOUB2021102}
J.~Ayoub, X.~J. Yang, and F.~Zhou, ``Modeling dispositional and initial learned
  trust in automated vehicles with predictability and explainability,''
  \emph{Transportation Research Part F: Traffic Psychology and Behaviour},
  vol.~77, pp. 102 -- 116, 2021.

\bibitem{weitz2019deep}
K.~Weitz, T.~Hassan, U.~Schmid, and J.-U. Garbas, ``Deep-learned faces of pain
  and emotions: Elucidating the differences of facial expressions with the help
  of explainable ai methods,'' \emph{tm-Technisches Messen}, vol.~86, no. 7-8,
  pp. 404--412, 2019.

\bibitem{elsner2013singular}
J.~B. Elsner and A.~A. Tsonis, \emph{Singular spectrum analysis: a new tool in
  time series analysis}.\hskip 1em plus 0.5em minus 0.4em\relax Springer
  Science \& Business Media, 2013.

\bibitem{lu2020entropy}
Y.~Lu, M.~Wang, W.~Wu, Q.~Zhang, Y.~Han, T.~Kausar, S.~Chen, M.~Liu, and
  B.~Wang, ``Entropy-based pattern learning based on singular spectrum analysis
  components for assessment of physiological signals,'' \emph{Complexity}, vol.
  2020, 2020.

\bibitem{rajesh2017classification}
K.~N. Rajesh and R.~Dhuli, ``Classification of ecg heartbeats using nonlinear
  decomposition methods and support vector machine,'' \emph{Computers in
  biology and medicine}, vol.~87, pp. 271--284, 2017.

\bibitem{lu2020dynamic}
Y.~Lu, M.~Wang, W.~Wu, Y.~Han, Q.~Zhang, and S.~Chen, ``Dynamic entropy-based
  pattern learning to identify emotions from eeg signals across individuals,''
  \emph{Measurement}, vol. 150, p. 107003, 2020.

\bibitem{kumar2014epileptic}
Y.~Kumar, M.~Dewal, and R.~Anand, ``Epileptic seizure detection using dwt based
  fuzzy approximate entropy and support vector machine,''
  \emph{Neurocomputing}, vol. 133, pp. 271--279, 2014.

\bibitem{jenke2014feature}
R.~Jenke, A.~Peer, and M.~Buss, ``Feature extraction and selection for emotion
  recognition from eeg,'' \emph{IEEE Transactions on Affective computing},
  vol.~5, no.~3, pp. 327--339, 2014.

\bibitem{chen2016xgboost}
T.~Chen and C.~Guestrin, ``Xgboost: A scalable tree boosting system,'' in
  \emph{Proceedings of the 22nd acm sigkdd international conference on
  knowledge discovery and data mining}, 2016, pp. 785--794.

\bibitem{ke2017lightgbm}
G.~Ke, Q.~Meng, T.~Finley, T.~Wang, W.~Chen, W.~Ma, Q.~Ye, and T.-Y. Liu,
  ``Lightgbm: A highly efficient gradient boosting decision tree,'' in
  \emph{Advances in neural information processing systems}, 2017, pp.
  3146--3154.

\bibitem{lundberg2020local}
S.~M. Lundberg, G.~Erion, H.~Chen, A.~DeGrave, J.~M. Prutkin, B.~Nair, R.~Katz,
  J.~Himmelfarb, N.~Bansal, and S.-I. Lee, ``From local explanations to global
  understanding with explainable ai for trees,'' \emph{Nature machine
  intelligence}, vol.~2, no.~1, pp. 2522--5839, 2020.

\bibitem{9652403}
K.~S. Kamble and J.~Sengupta, ``Ensemble machine learning-based affective
  computing for emotion recognition using dual-decomposed eeg signals,''
  \emph{IEEE Sensors Journal}, pp. 1--1, 2021.

\bibitem{zhou2022predicting}
F.~Zhou, A.~Alsaid, M.~Blommer, R.~Curry, R.~Swaminathan, D.~Kochhar,
  W.~Talamonti, and L.~Tijerina, ``Predicting driver fatigue in monotonous
  automated driving with explanation using gpboost and shap,''
  \emph{International Journal of Human--Computer Interaction}, vol.~38, no.~8,
  pp. 719--729, 2022.

\bibitem{zhou2021using}
F.~Zhou, X.~J. Yang, and J.~C. de~Winter, ``Using eye-tracking data to predict
  situation awareness in real time during takeover transitions in conditionally
  automated driving,'' \emph{IEEE Transactions on Intelligent Transportation
  Systems}, vol.~23, no.~3, pp. 2284--2295, 2021.

\bibitem{ayoub2021modeling}
J.~Ayoub, X.~J. Yang, and F.~Zhou, ``Modeling dispositional and initial learned
  trust in automated vehicles with predictability and explainability,''
  \emph{Transportation research part F: traffic psychology and behaviour},
  vol.~77, pp. 102--116, 2021.

\bibitem{shapley1953contributions}
L.~S. Shapley, H.~Kuhn, and A.~Tucker, ``Contributions to the theory of
  games,'' \emph{Annals of mathematics studies}, vol.~28, no.~2, pp. 307--317,
  1953.

\bibitem{lundberg2017unified}
S.~M. Lundberg and S.-I. Lee, ``A unified approach to interpreting model
  predictions,'' in \emph{Advances in neural information processing systems},
  2017, pp. 4765--4774.

\bibitem{yin2017recognition}
Z.~Yin, M.~Zhao, Y.~Wang, J.~Yang, and J.~Zhang, ``Recognition of emotions
  using multimodal physiological signals and an ensemble deep learning model,''
  \emph{Computer methods and programs in biomedicine}, vol. 140, pp. 93--110,
  2017.

\bibitem{picard2001toward}
R.~W. Picard, E.~Vyzas, and J.~Healey, ``Toward machine emotional intelligence:
  Analysis of affective physiological state,'' \emph{IEEE transactions on
  pattern analysis and machine intelligence}, vol.~23, no.~10, pp. 1175--1191,
  2001.

\bibitem{liu2008physiology}
C.~Liu, K.~Conn, N.~Sarkar, and W.~Stone, ``Physiology-based affect recognition
  for computer-assisted intervention of children with autism spectrum
  disorder,'' \emph{International journal of human-computer studies}, vol.~66,
  no.~9, pp. 662--677, 2008.

\bibitem{lin2017deep}
W.~Lin, C.~Li, and S.~Sun, ``Deep convolutional neural network for emotion
  recognition using eeg and peripheral physiological signal,'' in
  \emph{International Conference on Image and Graphics}.\hskip 1em plus 0.5em
  minus 0.4em\relax Springer, 2017, pp. 385--394.

\bibitem{nasoz2004emotion}
F.~Nasoz, K.~Alvarez, C.~L. Lisetti, and N.~Finkelstein, ``Emotion recognition
  from physiological signals using wireless sensors for presence
  technologies,'' \emph{Cognition, Technology \& Work}, vol.~6, no.~1, pp.
  4--14, 2004.

\bibitem{ringeval2015prediction}
F.~Ringeval, F.~Eyben, E.~Kroupi, A.~Yuce, J.-P. Thiran, T.~Ebrahimi,
  D.~Lalanne, and B.~Schuller, ``Prediction of asynchronous dimensional emotion
  ratings from audiovisual and physiological data,'' \emph{Pattern Recognition
  Letters}, vol.~66, pp. 22--30, 2015.

\bibitem{ringeval2013introducing}
F.~Ringeval, A.~Sonderegger, J.~Sauer, and D.~Lalanne, ``Introducing the recola
  multimodal corpus of remote collaborative and affective interactions,'' in
  \emph{2013 10th IEEE international conference and workshops on automatic face
  and gesture recognition (FG)}.\hskip 1em plus 0.5em minus 0.4em\relax IEEE,
  2013, pp. 1--8.

\bibitem{li2016analysis}
C.~Li, C.~Xu, and Z.~Feng, ``Analysis of physiological for emotion recognition
  with the irs model,'' \emph{Neurocomputing}, vol. 178, pp. 103--111, 2016.

\bibitem{hassan2019human}
M.~M. Hassan, M.~G.~R. Alam, M.~Z. Uddin, S.~Huda, A.~Almogren, and G.~Fortino,
  ``Human emotion recognition using deep belief network architecture,''
  \emph{Information Fusion}, vol.~51, pp. 10--18, 2019.

\bibitem{dehghani2019subject}
A.~Dehghani, T.~Glatard, and E.~Shihab, ``Subject cross validation in human
  activity recognition,'' \emph{arXiv preprint arXiv:1904.02666}, 2019.

\bibitem{koul2018cross}
A.~Koul, C.~Becchio, and A.~Cavallo, ``Cross-validation approaches for
  replicability in psychology,'' \emph{Frontiers in Psychology}, vol.~9, p.
  1117, 2018.

\bibitem{saeb2017need}
S.~Saeb, L.~Lonini, A.~Jayaraman, D.~C. Mohr, and K.~P. Kording, ``The need to
  approximate the use-case in clinical machine learning,'' \emph{Gigascience},
  vol.~6, no.~5, p. gix019, 2017.

\bibitem{kandemir2014multi}
M.~Kandemir, A.~Vetek, M.~Goenen, A.~Klami, and S.~Kaski, ``Multi-task and
  multi-view learning of user state,'' \emph{Neurocomputing}, vol. 139, pp.
  97--106, 2014.

\bibitem{zhong2017emotion}
B.~Zhong, Z.~Qin, S.~Yang, J.~Chen, N.~Mudrick, M.~Taub, R.~Azevedo, and
  E.~Lobaton, ``Emotion recognition with facial expressions and physiological
  signals,'' in \emph{2017 IEEE Symposium Series on Computational Intelligence
  (SSCI)}.\hskip 1em plus 0.5em minus 0.4em\relax IEEE, 2017, pp. 1--8.

\bibitem{gupta2019cross}
V.~Gupta, M.~D. Chopda, and R.~B. Pachori, ``Cross-subject emotion recognition
  using flexible analytic wavelet transform from eeg signals,'' \emph{IEEE
  Sensors Journal}, vol.~19, no.~6, pp. 2266--2274, 2019.

\bibitem{ayoub2021combat}
J.~Ayoub, X.~J. Yang, and F.~Zhou, ``Combat covid-19 infodemic using
  explainable natural language processing models,'' \emph{Information
  Processing \& Management}, vol.~58, no.~4, p. 102569, 2021.

\bibitem{murdoch2019definitions}
W.~J. Murdoch, C.~Singh, K.~Kumbier, R.~Abbasi-Asl, and B.~Yu, ``Definitions,
  methods, and applications in interpretable machine learning,''
  \emph{Proceedings of the National Academy of Sciences}, vol. 116, no.~44, pp.
  22\,071--22\,080, 2019.

\bibitem{piho2018mutual}
L.~Piho and T.~Tjahjadi, ``A mutual information based adaptive windowing of
  informative eeg for emotion recognition,'' \emph{IEEE Transactions on
  Affective Computing}, 2018.

\bibitem{benedek2010decomposition}
M.~Benedek and C.~Kaernbach, ``Decomposition of skin conductance data by means
  of nonnegative deconvolution,'' \emph{Psychophysiology}, vol.~47, no.~4, pp.
  647--658, 2010.

\bibitem{pincus1991approximate}
S.~M. Pincus, ``Approximate entropy as a measure of system complexity.''
  \emph{Proceedings of the National Academy of Sciences}, vol.~88, no.~6, pp.
  2297--2301, 1991.

\bibitem{richman2000physiological}
J.~S. Richman and J.~R. Moorman, ``Physiological time-series analysis using
  approximate entropy and sample entropy,'' \emph{American Journal of
  Physiology-Heart and Circulatory Physiology}, vol. 278, no.~6, pp.
  H2039--H2049, 2000.

\bibitem{chen2007characterization}
W.~Chen, Z.~Wang, H.~Xie, and W.~Yu, ``Characterization of surface emg signal
  based on fuzzy entropy,'' \emph{IEEE Transactions on neural systems and
  rehabilitation engineering}, vol.~15, no.~2, pp. 266--272, 2007.

\bibitem{friedman2001greedy}
J.~H. Friedman, ``Greedy function approximation: a gradient boosting machine,''
  \emph{Annals of statistics}, pp. 1189--1232, 2001.

\bibitem{tang2017multimodal}
H.~Tang, W.~Liu, W.-L. Zheng, and B.-L. Lu, ``Multimodal emotion recognition
  using deep neural networks,'' in \emph{International Conference on Neural
  Information Processing}.\hskip 1em plus 0.5em minus 0.4em\relax Springer,
  2017, pp. 811--819.

\bibitem{alonso2005application}
F.~Alonso, J.~Del~Castillo, and P.~Pintado, ``Application of singular spectrum
  analysis to the smoothing of raw kinematic signals,'' \emph{Journal of
  biomechanics}, vol.~38, no.~5, pp. 1085--1092, 2005.

\bibitem{delgado2019approximate}
A.~Delgado-Bonal and A.~Marshak, ``Approximate entropy and sample entropy: A
  comprehensive tutorial,'' \emph{Entropy}, vol.~21, no.~6, p. 541, 2019.

\bibitem{SUBASI201927}
A.~Subasi, ``Chapter 2 - biomedical signals,'' in \emph{Practical Guide for
  Biomedical Signals Analysis Using Machine Learning Techniques}, A.~Subasi,
  Ed.\hskip 1em plus 0.5em minus 0.4em\relax Academic Press, 2019, pp. 27 --
  87.

\bibitem{doukas2008intelligent}
C.~Doukas and I.~Maglogiannis, ``Intelligent pervasive healthcare systems,'' in
  \emph{Advanced Computational Intelligence Paradigms in Healthcare-3}.\hskip
  1em plus 0.5em minus 0.4em\relax Springer, 2008, pp. 95--115.

\end{thebibliography}

\begin{IEEEbiography}[{\includegraphics[width=1in,height=1.55in,clip,keepaspectratio]{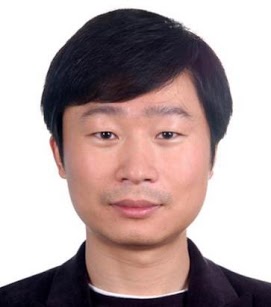}}]{Dr. Feng Zhou received the Ph.D. degree in Human Factors Engineering from Nanyang Technological University, Singapore, in 2011 and Ph.D. degree in Mechanical Engineering from Gatech in 2014. He was a Research Scientist at MediaScience, Austin TX, from 2015 to 2017. He is currently an Assistant Professor with the Department of Industrial and Manufacturing Systems Engineering, University of Michigan, Dearborn. His main research interests include human factors, human-computer interaction, engineering design, and human-centered design.}
\end{IEEEbiography}
\begin{IEEEbiography}[{\includegraphics[width=1in,height=1.55in,clip,keepaspectratio]{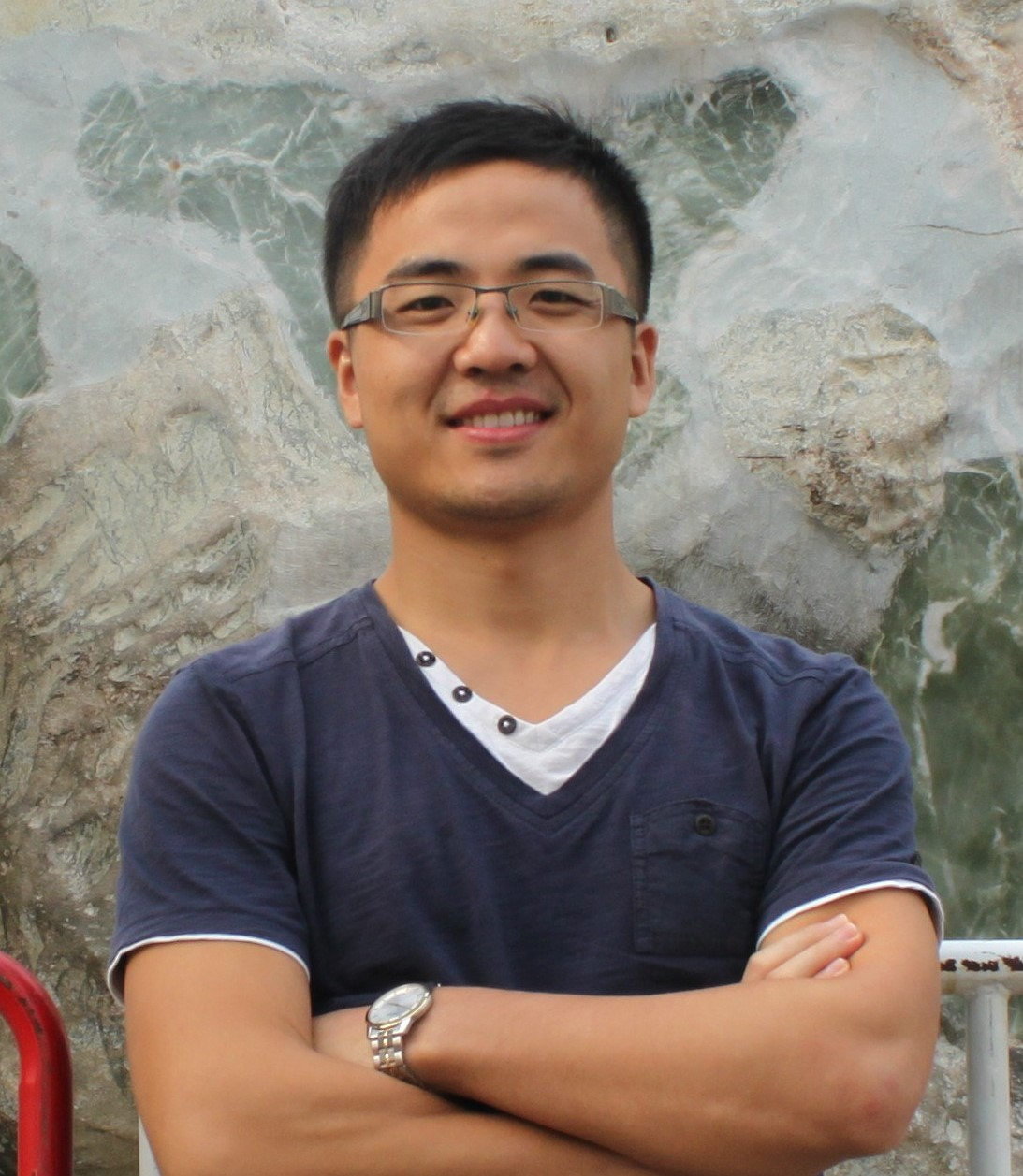}}]{Dr. Tao Chen received the Ph.D. degree in information engineering from Nanyang Technological University, Singapore, in 2013. He was a Research Scientist at the Institute for Infocomm Research, A*STAR, Singapore, from 2013 to 2017, and a Senior Scientist at the Huawei Singapore Research Center from 2017 to 2018. He is currently a Professor with the School of Information Science and Technology, Fudan University, Shanghai, China. His main research interests include computer vision and machine learning.}
\end{IEEEbiography}
\begin{IEEEbiography}[{\includegraphics[width=1in,height=1.55in,clip,keepaspectratio]{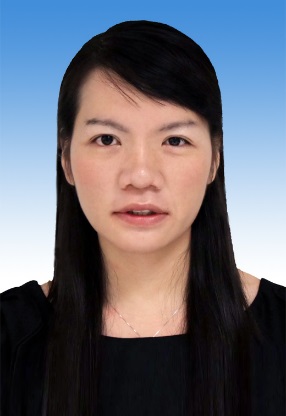}}]{Dr.Baiying Lei (Senior Member, IEEE) received the Ph.D. degree from Nanyang Technological University (NTU), Singapore, in 2013. She is currently a distinguised professor with the Health Science Center, School of Biomedical Engineering, Shenzhen University, Shenzhen, China. She has coauthored more than 220 scientific articles. Her research interests include medical image analysis, machine learning, and pattern recognition. Prof. Lei serves as an Associate Editor for the IEEE TMI and IEEE TNNLS, and an Editorial Board Member of Medical Image Analysis, Neural Computing and Application.}
\end{IEEEbiography}




\end{document}